\documentclass[superscriptaddress,amsmath,amssymb,twocolumn]{revtex4-1}

\pdfoutput=1

\usepackage{graphicx}
\usepackage{chemarr,textcomp}
\usepackage{times,longtable}
\usepackage{hyperref}
\usepackage{color,multirow}
\usepackage{afterpage,enumitem}
\usepackage{soul,comment}
\usepackage[usenames,dvipsnames]{xcolor}
\usepackage{colortbl}
\usepackage{pbox}
\usepackage{capt-of}% or use the larger `caption` package%
\graphicspath{{large/}}
\bibpunct{[}{]}{,}{n}{}{,}

\newcommand{\ovl}[1]{\overline{#1}}
\newcommand{\udl}[1]{\underline{#1}}
\newcommand{\hatt}[1]{\widehat{#1}}

\definecolor{n1}{rgb}{1,0.7,0}
\definecolor{n23}{rgb}{0.8,0,0}
\definecolor{n45}{rgb}{0,0,0.8}
\definecolor{n6}{rgb}{.2,0.75,0.2}
\definecolor{n7}{rgb}{0.3,0.8,0.3}

\definecolor{m1}{rgb}{1,0,0}
\definecolor{m2}{rgb}{.8,0,.8}
\definecolor{m3}{rgb}{.3,0,1}
\definecolor{m4}{rgb}{.8,0.5,0}
\definecolor{m5}{rgb}{.5,.8,0}

\newcommand{\avg}[1]{\left\langle #1 \right\rangle}
\renewcommand{\bar}[1]{\overline{#1}}

\renewcommand{\hat}[1]{\widehat{#1}}

\newcommand{\ii}{\mathrm{i}}

\usepackage{chemarr}
\usepackage{epsf,mathtools}
\usepackage{graphicx}% Include figure files
\usepackage{dcolumn}% Align table columns on decimal point
\usepackage{bm}% bold math
\graphicspath{ {imagespdf/} }

\begin{document}

%\title{Competition and effective RNA-RNA interactions in minimal models of Post-Transcriptional Regulatory Networks}

\title{Path-integral solution of MacArthur's resource-competition model for large ecosystems with random species-resources couplings}

\author{A.R. Batista-Tom\'as}
\affiliation{Group of Complex Systems and Statistical Physics,  Department of Applied Physics, Physics Faculty, University of Havana, 10400 Cuba}

\author{Andrea De Martino}
\email{Corresponding author. Email: andrea.demartino@roma1.infn.it}
\affiliation{Soft and Living Matter Lab, Institute of Nanotechnology (CNR-NANOTEC), Consiglio Nazionale delle Ricerche, Rome 00185, Italy}
\affiliation{Italian Institute for Genomic Medicine, Turin 10060, Italy}

\author{Roberto Mulet}
\affiliation{Group of Complex Systems and Statistical Physics,  Department of Applied Physics, Physics Faculty, University of Havana, 10400 Cuba}

\begin{abstract}
We solve MacArthur's resource-competition model with random species-resource couplings in the `thermodynamic' limit of infinitely many species and resources using dynamical path-integrals \`a la De Domincis. We analyze how the steady state picture changes upon modifying several parameters, including the degree of heterogeneity of metabolic strategies (encoding the preferences of species) and of maximal resource levels (carrying capacities), and discuss its stability. Ultimately, the scenario obtained by other approaches is recovered by analyzing an effective one-species-one-resource ecosystem that is fully equivalent to the original multi-species one. The technique used here can be applied for the analysis of other model ecosystems related to the version of MacArthur's model considered here.
\end{abstract}

\maketitle

{\bf Mathematical models of ecosystems have repeatedly proved useful to understand how species survival, global stability and responses to perturbations are controlled by the parameters governing the interactions between species and resources. Following Wigner's recipe, large ecosystems with extended, complicated and unknown interaction networks can be usefully modeled by assuming quenched random species-resources couplings. In such cases, the statistical mechanics of disordered systems provides tools to calculate macroscopic properties as averages over network realizations. Here we apply one such method, based on dynamical generating functionals, to MacArthur's resource competition model, a classic model of an ecosystem in which different species compete for a pool of resources.}

\section{Introduction}

Despite the complicated structure and dynamics of underlying interactions, microbial ecosystems generate robust statistical outcomes that can now be experimentally probed at genomic resolution \cite{hekstra12,huttenhower12,friedman17,golford18}. To understand the origins and richness of these features and to derive testable predictions, a variety of mathematical models have been considered over time, ranging from coarse-grained ones based on simple differential equations to metabolically-realistic schemes capable of accounting for time- and space-dependencies  \cite{harcombe14,wade16,succurro18,biroli18,goyal18}. Statistical mechanics approaches are particularly well suited to analyze large instances of such models and derive macroscopic laws, especially when the complex web of intraspecific and trophic interactions can be modeled by quenched random variables.

Among the schemes that have attracted attention, MacArthur's resource competition model plays a central role as a theoretical reference frame in view of its flexibility, phenomenological richness and the possibility of studying highly diverse communities with quantitative detail \cite{macarthur70,chesson90}. The emergent statistical properties of several versions of MacArthur's model have recently been studied by approaches rooted in the theory of disordered systems, including replica theory and the cavity method (see e.g. \cite{demartino06,yoshino07,tikhonov16,tikhonov16-2,tikhonov17,advani18,tikhonov18,cui20,marsland20,pacciani20}), to address features as diverse as the role of heterogeneity of species and resources for the stability of large ecosystems \cite{demartino06,yoshino07}, the emergence of robust population structures \cite{tikhonov16-2,marsland20}, the impact of different kinds of resource supply dynamics \cite{cui20}, or the success of different eco-evolutionary strategies of invasion in small versus large ecosystems \cite{tikhonov18}. 

In this work we contribute to this line of studies by solving MacArthur's model using dynamical mean field theory (DMFT) \cite{dedo78,sompo82}. The key advantage of this method over  replicas lies in the fact that it does not rely on specific properties of interactions (e.g. symmetry) or on the existence of a Lyapunov function of the dynamics. On the other hand, dynamical path integrals allow to treat such cases in a more straightforward manner compared to the cavity method. This makes DMFT more broadly applicable, at least in principle. In addition, DMFT reduces the dynamics of the multi-species, multi-resources system to a pair of processes involving one effective species and one effective resource, whose structure renders explicit the role of parameters that only impact the original dynamics implicitly.

Besides retrieving the picture derived by other means, more precisely by the cavity method \cite{advani18}, we will analyze the stability properties, the role of resource heterogeneity for the survival and numerosity of species, and how a species' survival probability is modulated by its prior fitness (i.e. by the fitness it would face in that environment in absence of other competing species).

\section{MacArthur's model}

For this paper, MacArthur's model is defined as e.g. in \cite{advani18}. We consider a system in which $N$ species interact through $M$ shared resources. If $n_i$ denotes the population size of species $i$ ($i=1,\ldots,N$) and $r^\mu$ denotes the level of resource $\mu$ ($\mu=1,\ldots,M$), its time evolution is described by the coupled equations for their respective growth rates $g_i$ and $g^\mu$ ($\dot{x}=\frac{dx}{dt}$)
    \begin{gather}\label{ndot}
    g_i\equiv\frac{\dot n_i}{n_i}= \sum_\mu c_i^\mu r^\mu-m_i~~,\\
    g^\mu\equiv\frac{\dot r^\mu}{r^\mu}=a^\mu(K^\mu-r^\mu)-\sum_j c_j^\mu n_j~~,\label{rdot}
    \end{gather}
where $c_i^\mu$ measures the contribution of resource $\mu$ to the growth rate (fitness) of species $i$, $m_i$ is a maintenance level for species $i$ (growth only occurs if the benefit from resources exceeds this threshold), $a^\mu$ is the intrinsic growth rate of resource $\mu$ and $K^\mu$ is the carrying capacity of resource $\mu$. In short, in such a setting each resource is externally supplied, so that in absence of species it undergoes logistic growth with rate $a^\mu$ and saturation level $K^\mu$. When species are present, they predate on resources with fitness benefits described by the coefficients $c_i^\mu$. 

Some basic properties of MacArthur's model are easily derived from (\ref{ndot}) and (\ref{rdot}) under the simplifying assumption that resources equilibrate much faster than populations. Indeed in this case one can replace $r^\mu$ in (\ref{ndot}) by its steady state value obtained by imposing $g^\mu=0$. Eq. (\ref{ndot}) can then be re-cast as
\begin{gather}\label{simpl}
g_i= -\frac{\partial L}{\partial n_i}~~,
\end{gather}
with 
\begin{gather}\label{lyap}
L=\frac{1}{2}\sum_\mu{}^{'}\frac{1}{a^\mu}\left(\sum_j n_j c_j^\mu-K^\mu a^\mu\right)^2+\sum_i m_i n_i~~,
\end{gather}
where the prime indicates that the sum runs over resources whose level is not zero (`non-depleted' for short) at stationarity. As $\dot{L}=-\sum_i (\dot n_i)^2/n_i\leq 0$, $L$ is a (convex) Lyapunov function of the dynamics and its unique minimum describes the steady state population sizes (and in turn resource levels). At stationarity, in particular, the population sizes of surviving species obey the conditions
\begin{gather}\label{s-s}
\sum_j J_{ij} n_j+h_i=0~~,
\end{gather}
where $J_{ij}=\sum_\mu' c_i^\mu c_j^\mu/a^\mu$ and $h_i=m_i-\sum_\mu' c_i^\mu K^\mu$. Because the rank of the matrix with elements $J_{ij}$ is at most equal to the number $M_s$ of non-depleted resources, the system (\ref{s-s}) contains at most $M_s$ independent equations. The number $N_s$ of surviving species (i.e. of variables in (\ref{s-s})) therefore satisfies $N_s\leq M_s$. This implies that the fractions $\phi_s$ and $\psi_s$ of surviving species and resources, respectively, are related by
\begin{gather}
\phi_s\leq\frac{\psi_s}{\nu}~~~~~,~~\nu=\frac{N}{M}~~. \label{nu}
\end{gather}
We shall see that this bound provides a quantitatively accurate description of the relationship between $\phi_s$ and $\psi_s$ even in the general case in which the timescales of (\ref{ndot}) and (\ref{rdot}) are not widely separate.

To model complex interdependencies of species on resources in extended ecosystems, it is normally assumed that $N$ and $M$ are large and that $c_i^\mu$s are quenched iid random variables. More specifically, the statistical mechanics approach studies the statistical properties emerging when $N,M\to\infty$ at fixed $\nu$ (see (\ref{nu})) and $c_i^\mu$s are quenched iid random variables with mean $c/N$ and variance $\sigma_c^2/N$. $\sigma_c^2$ is an especially important parameter here, as it provides a proxy for the metabolic heterogeneity of species. In the linear approximation and within the same assumptions leading to (\ref{nu}), setting $a^\mu=1$ for simplicity, one easily sees that a perturbation $\delta n_i$ of a steady state population $\bar n_i>0$ for the MacArthur model evolves as
\begin{gather}
\dot{\delta n_i}=-\bar n_i\sum_j \mathbf{K}_{ij}\delta n_j~~~~~,~~~\mathbf{K}_{ij}=\sum_\mu{}^{'} c_i^\mu c_j^\mu~~.
\end{gather}
Stability for the above system is governed by the smallest eigenvalue of the random matrix $\{\mathbf{K}_{ij}\}$, which can be calculated using results from \cite{sengupta99}. One finds 
\begin{gather}\label{lmin}
\lambda_{\min}=\sigma_c\left(\sqrt{\frac{\psi_s}{\nu}}-\sqrt{\phi_s}\right)~~.
\end{gather}
This formula can provide qualitative information about the linear stability of the ecosystem's dynamics. In agreement with (\ref{nu}), $\lambda_{\min}$ vanishes when $\nu=\psi_s/\phi_s$. Moreover, as both $\phi_s$ and $\psi_s$ are bound to change with $\sigma_c$, (\ref{lmin}) implies the possibility of a non-trivial dependence of $\lambda_{\min}$ on species heterogeneity. This issue will be explored more thoroughly in what follows.

Besides that encoded in the coefficients $c_i^\mu$, other sources of heterogeneity can be accounted for via randomness in other parameters. For instance, to model diverse availability of resources $K^\mu$s can be taken to be quenched iid random variables with mean $K$ and variance $\sigma^2_K$. An important role in our results will be played  by the `prior fitness' of species $i$, defined as
\begin{gather}\label{prior}
    f_{i,0}\equiv \sum_{\mu=1}^M c_i^\mu K^\mu-m_i~~,
\end{gather}
in terms of which (\ref{ndot}) takes the form
\begin{gather}\label{ndot2}
g_i= f_{i,0}-\sum_\mu c_i^\mu (K^\mu-r^\mu)~~.
\end{gather}
This clarifies how the participation in resource competition affects the {\it a priori} viability of species $i$. $f_{i,0}$ measures the  fitness that a species would have if it was introduced in the environment at maximal resource levels and in absence of other species. It is reasonable to expect that $f_{i,0}$ correlates positively with the survival probability also in a complex ecosystem with $N$ species. This was shown numerically in \cite{advani18}. We shall further explore this issue here by deriving an approximate quantitative relationship linking these quantities.

\section{Dynamical generating function and effective processes}

The path integral approach to the system formed by (\ref{ndot}) and (\ref{rdot}) is based on the computation of the dynamical generating function 
\begin{equation}
Z[\boldsymbol{\psi,\phi}]\equiv\ovl{\avg{e^{\ii\sum_i\int dt \,\psi_i(t)n_i(t)+\ii\sum_\mu\int dt \,\phi^\mu(t)r^\mu(t)}}_\mathrm{paths}}~~,\label{paths}
\end{equation}
where the quantities $\boldsymbol{\psi}=\{\psi_i\}$ and $\boldsymbol{\phi}=\{\phi^\mu\}$ represent auxiliary fields associated respectively to the variables $\mathbf{n}=\{n_i\}$ and $\mathbf{r}=\{r^\mu\}$, the over-bar denotes an average over quenched disorder (most notably the coefficients $c_i^\mu$), while the brackets denote an average over the realizations (`paths') of the dynamics of the system at fixed disorder, which includes possibly random initial conditions. The key advantage of $Z$ lies in the fact that the moments of the variables $n_i$ and $r^\mu$ can be written in terms of derivatives of $Z$ over the auxiliary fields. For instance, the mean values of $n_i$ and $r^\mu$ (over paths and disorder) are given by
\begin{gather}
\ovl{\avg{n_i(t)}_\mathrm{paths}}=-\ii\lim_{\substack{\boldsymbol{\psi}\to 0\\ \boldsymbol{\phi}\to 0}}\frac{\partial Z}{\partial \psi_i(t)}~~,\\
\ovl{\avg{r^\mu(t)}_\mathrm{paths}}=-\ii\lim_{\substack{\boldsymbol{\psi}\to 0\\ \boldsymbol{\phi}\to 0}}\frac{\partial Z}{\partial \phi^\mu(t)}~~.
\end{gather}
Notice that these expressions are valid at any time $t$. As shown in the Appendix, the explicit calculation of the average over disorder leads to the identification of a set of macroscopic order parameters whose averages provide a full characterization of the $N$-species, $M$-resources system. These parameters include the mean population sizes and resource levels, i.e.
\begin{gather}
\rho_n(t)=\frac{1}{N}\sum_{i=1}^N n_i(t)~~,\\
\rho_r(t)=\frac{1}{N}\sum_{\mu=1}^M r^\mu(t)~~,\label{ror}
\end{gather}
and the two-time correlation functions
\begin{gather}
Q_n(t,t')=\frac{1}{N}\sum_{i=1}^N n_i(t)n_i(t')~~,\\
Q_r(t,t')=\frac{1}{N}\sum_{\mu=1}^M r^\mu(t)r^\mu(t')~~.
\end{gather} 
In view of the connection between $Z$ and moments of $n_i$ and $r^\mu$, it is clear that $Z$ allows for a full statistical description of the dynamics.

Ultimately, $Z$ can be evaluated in the limit $N,M\to\infty$ (with fixed $\nu=N/M$) via a saddle-point method following which the full dynamics is re-cast in terms of a pair of effective stochastic processes, one for an effective species with population size $n(t)$, the other for an effective resource with level $r(t)$ (see Appendix for the derivation). Such processes read
\begin{widetext}
\begin{gather}
\frac{\dot{n}(t)}{n(t)}=c\rho_r(t)-m-\frac{\sigma_c^2}{\nu} \int dt' \,G_r(t,t')n(t')+\theta(t)+\xi_n(t)~~,%~,~~~\avg{\xi_n(t)\xi_n(t')}=\sigma_c^2 Q_r(t,t')~~,
\label{effn}\\
\frac{\dot{r}(t)}{r(t)}=a(K-r(t))-c\rho_n(t)-\sigma_c^2\int dt' \,G_n(t,t')r(t')+\eta(t)+\xi_r(t)~~,%~,~~~\avg{\xi_r(t)\xi_r(t')}=\sigma_c^2 Q_n(t,t')+a^2\sigma^2_K~~,
\label{effr}
\end{gather}
\end{widetext}
where $m$ is the maintenance level of the effective species, $a$ and $K$ denote respectively the rate of growth and the carrying capacity of the effective resource, $\xi_n$ and $\xi_r$ are zero-average Gaussian random variables with
\begin{gather}
\avg{\xi_n(t)\xi_n(t')}=\sigma_c^2 Q_r(t,t')~~,\\
\avg{\xi_r(t)\xi_r(t')}=\sigma_c^2 Q_n(t,t')+a^2\sigma^2_K~~,
\end{gather}
$\theta(t)$ and $\eta(t)$ are auxiliary probing fields, while $G_r$ and $G_n$ are interaction kernels describing the response functions, i.e. 
\begin{gather}
G_n(t,t')=\avg{\frac{\partial n(t)}{\partial\theta(t')}}_\star~~,\\
G_r(t,t')=\avg{\frac{\partial r(t)}{\partial\eta(t')}}_\star~~,
\end{gather}
with $\avg{\cdot}_\star$ denoting an average over realizations of (\ref{effn},\ref{effr}). 

The above equations clarify how disorder affects the dynamics in the limit of large ecosystems. Heterogeneity in metabolic strategies (i.e. $\sigma_c^2>0$) translates into a term that couples species and resources along the entire dynamical trajectory. Hence, quenched random $c_i^\mu$s generate long-term memory in the dynamics despite the fact that the model defined by (\ref{ndot}) and (\ref{rdot}) is Markovian. For $\sigma_K^2=0$ and large enough $\sigma_c^2$, moreover, the memory term dominates over the noise term, as the strength of the latter is proportional to $\sigma_c$. On the other hand, heterogeneity in carrying capacities (i.e. $\sigma_K^2$) only affects the strength of the noise term in (\ref{effr}). Therefore, when the resource dynamics is dominated by noise in $K$s, we should expect to see that resource levels change in an effectively random way. As we shall see below in detail, such a variability bears a non-trivial impact on the abundance of species at steady state. Notice that, as $\sigma_c$ increases, negative values of $c_i^\mu$, corresponding to resources having a detrimental effect on species $i$ (e.g. toxic compounds), will become more frequent. Likewise, as $\sigma_K$ increases some resources will have a `negative carrying capacity', corresponding in effect to a sink that drains those resources from the environment.

It is instructive to study the behaviour of (\ref{effn}) at small times. Noting that $G_r(0,0)=0$ by causality and that $Q_r(0,0)=d$ (constant), one sees that the quantity $g_n(0)=\dot n(0)/n(0)$ is a Gaussian variable with mean $c\rho_r(0)-m$ and variance $d\sigma^2_c$. Moreover, by virtue of (\ref{ror}) one has $\rho_r(0)=r_0/\nu$, where $r_0$ denotes the mean of $r(0)$ over the distribution of initial conditions of (\ref{effr}). Therefore, on average, the effective species initially increases its population if
\begin{gather}
r_0>\frac{m\nu}{c}~~.
\end{gather}
As might have been expected, the initial growth rate is more likely positive in less competitive ecosystems, for species with less demanding maintenance requirements and/or when metabolic strategies are more efficient. More precisely, the probability that $g_n(0)>0$ is given by $\pi_0\equiv\mathrm{Prob}\left\{\xi_n(0)>m-\frac{cr_0}{\nu}\right\}$, i.e.
\begin{gather}
\pi_0=\frac{1}{2}\,\mathrm{erfc}\,\frac{\nu m-cr_0}{\nu\sigma_c\sqrt{2d}}~~.
\end{gather}
Hence in highly competitive ecosystems ($\nu\gg 1$), $\pi_0\simeq\frac{1}{2}\,\mathrm{erfc}\,\frac{m}{\sigma_c\sqrt{2d}}$, implying that a stronger metabolic heterogeneity yields a higher initial fitness.

\section{Stationary state}

\begin{figure}
 \includegraphics[width=.4\textwidth]{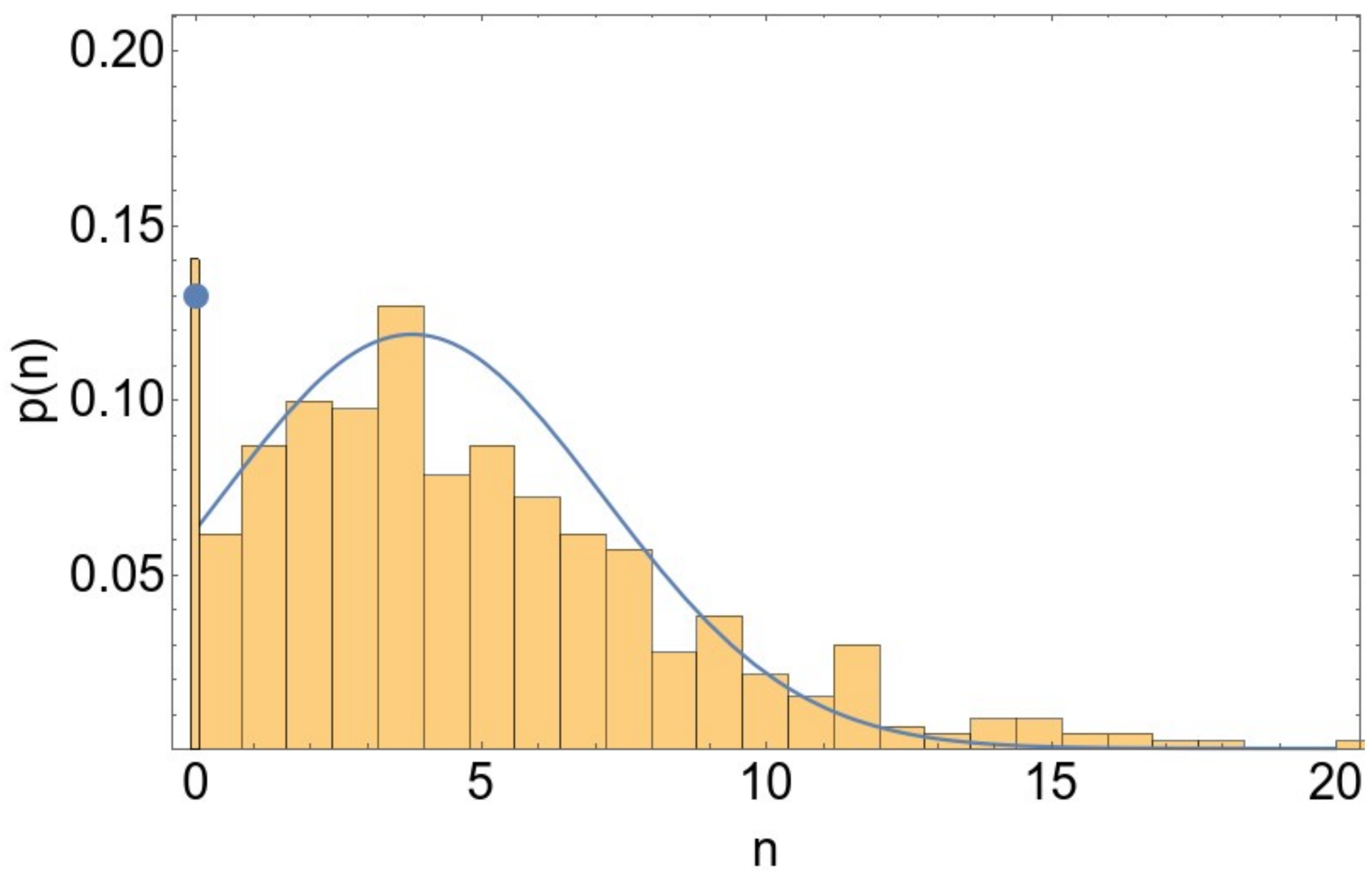}
 \label{fig:hist}
 \caption{\label{distr} Representative example of the distribution $p(n)$ of population sizes for $\nu=0.22,K=5,\sigma^2_c=a=m=1$ and $\sigma^2_K=0$. The solid line is the prediction obtained from (\ref{distr_th}), the blue marker denoting the weight of the peak at $n=0$ (i.e. the fraction of extinct species). Bars represent instead the histogram of the population sizes of $10^3$ species obtained by averaging the steady state populations of $100$ independent realizations of  MacArthur's model (\ref{ndot}) and (\ref{rdot}).}
\end{figure}

In the steady state defined by $\dot{n}=\dot{r}=0$, we have $n(t)\to \udl{n}$ and $r(t)\to \udl{r}$ while two-time quantities are bound to be time-translation invariant, e.g.
\begin{gather}
\lim_{t\to\infty} Q_n(t+\tau,\tau)= Q_n(\tau)~~,
\end{gather}
with finite means, average correlations and integrated responses, i.e. ($x=n,r$)
\begin{gather}
\lim_{\tau\to\infty}\frac{1}{\tau}\int_0^\tau dt\,\rho_x(t)=\bar\rho_x~~,\\
\lim_{\tau\to\infty}\frac{1}{\tau}\int_0^\tau dt\, Q_x(t)=q_x~~,\\
\lim_{\tau\to\infty}\int_0^\tau dt\,G_x(t)=\chi_x~~.
\end{gather}
Using these properties and definitions, one  shows, for instance, that
\begin{multline}
\lim_{\tau\to\infty}\frac{1}{\tau}\int_0^\tau dt\int_0^t dt'\,G_r(t-t')n(t')\\
=\int_0^\infty ds\, G_r(s)\lim_{\tau\to\infty}\frac{1}{\tau}\int_s^\tau dt\, n(t-s)\\
=\int_0^\infty ds\, G_r(s)\lim_{\tau\to\infty}\frac{1}{\tau}\int_0^{\tau-s} ds'\, n(s')\\
=\int_0^\infty ds\, G_r(s)\lim_{\tau\to\infty}\frac{\tau-s}{\tau}\frac{1}{\tau-s}\int_0^{\tau-s} ds'\, n(s')\\
=\chi_r\udl{n}~~.
\end{multline}
Hence at stationarity (\ref{effn}) and (\ref{effr}) are easily seen to imply 
\begin{gather}
c\bar\rho_r-\frac{\sigma_c^2}{\nu}\chi_r \udl{n}+\sigma_c\sqrt{q_r}\,z=m~~,\\
\udl{r}\left(a+\sigma_c^2\chi_n\right)=aK-c\bar\rho_n+\sqrt{\sigma^2_cq_n+a^2\sigma^2_K}\,z~~,
\end{gather}
with $z$ a Gaussian random variable with zero mean and unit variance. (The probing fields $\theta$ and $\eta$ can be set to zero, noting that differentiating with respect to them is equivalent to differentiating with respect to $z$, modulo constant factors.) The above equations coincide with those obtained in \cite{advani18} by the cavity method. Since $\udl{n},\udl{r}\geq 0$, the steady state population size and resource level are given by
\begin{gather}\label{26}
\udl{n}\equiv \udl{n}(z)=\frac{c\bar\rho_r-m+\sigma_c\sqrt{q_r}z}{\sigma_c^2\chi_r/\nu}\,\Theta(z-z_n)~~,\\
\udl{r}\equiv \udl{r}(z)=\frac{aK-c\bar\rho_n+\sqrt{\sigma^2_cq_n+a^2\sigma^2_K}\;z}{a+\sigma_c^2\chi_n}\,\Theta(z-z_r)~~,\label{27}
\end{gather}
where $\Theta(x)$ is the step function while
\begin{gather}
z_n\equiv\frac{m-c\bar\rho_r}{\sigma_c\sqrt{q_r}}~~,\\ z_r\equiv\frac{c\bar\rho_n-aK}{\sqrt{\sigma^2_cq_n+a^2\sigma^2_K}}\label{zr}~~.
\end{gather}
Equations (\ref{26}) and (\ref{27}) finally allow to compute macroscopic parameters from 
\begin{gather}
\bar\rho_n\equiv\avg{n}_\star=\avg{\udl{n}}_z \label{prima}\\
\bar\rho_r\equiv\frac{1}{\nu}\avg{r}_\star=\frac{1}{\nu}\avg{\udl{r}}_z\\
q_n\equiv\avg{n^2}_\star=\avg{\udl{n}^2}_z\\
q_r\equiv\frac{1}{\nu}\avg{r^2}_\star=\frac{1}{\nu}\avg{\udl{r}^2}_z\\
\chi_n=\frac{1}{\sigma_c\sqrt{q_r}}\avg{\frac{\partial\udl{n}}{\partial z}}_z\\
\chi_r=\frac{1}{\sigma_c\sqrt{q_n}}\avg{\frac{\partial\udl{r}}{\partial z}}_z\label{ultima}
\end{gather}
After some algebra, one ends up with the set of equations
\begin{gather}
\bar \rho_n= \frac{\nu \sqrt{q_r}}{\sigma_c\chi_r}\left[\frac{e^{-\frac{z_n^2}{2}}}{\sqrt{2\pi}}-\frac{z_n}{2}\,\text{erfc}\left(\frac{z_n}{\sqrt 2}\right)\right]\\
\bar \rho_r= \frac{\sqrt{\sigma^2_cq_n+a^2\sigma^2_K}}{\nu(a+\sigma_c^2 \chi_n) }\left[\frac{e^{-\frac{z_r^2}{2}}}{\sqrt{2\pi}}-\frac{z_r}{2}\,\text{erfc}\left(\frac{z_r}{\sqrt 2}\right)\right]\\
q_n=\frac{\nu^2 q_r}{\sigma_c^2\chi_r^2}\left[\frac{1}{2}(1+z_n^2)\,\text{erfc}\left(\frac{z_n}{\sqrt 2}\right)-\frac{z_n e^{-\frac{z_n^2}{2}}}{\sqrt{2\pi}}\right]\\
q_r=\frac{\sigma_c^2 q_n+a^2\sigma_K^2}{\nu(a+\sigma_c^2\chi_n)^2}\left[\frac{1}{2}(1+z_r^2)\,\text{erfc}\left(\frac{z_r}{\sqrt 2}\right)-\frac{z_r e^{-\frac{z_r^2}{2}}}{\sqrt{2\pi}}\right]\\
\chi_n=\frac{\nu}{2\sigma_c^2\chi_r}\,\text{erfc}\left(\frac{z_n}{\sqrt 2}\right)\\
\chi_r=\frac{1}{2(a+\sigma_c^2\chi_n)}\,\text{erfc}\left(\frac{z_r}{\sqrt 2}\right)
\end{gather}
which can be solved numerically for any choice of the parameters $\nu$, $c$, $a$, $\sigma_c$ and $\sigma_K$. In turn, one can obtain the fractions of surviving species and non-depleted resources from
\begin{gather}
\phi_s=\frac{1}{2}\,\text{erfc}\left(\frac{z_n}{\sqrt 2}\right)~~,\\
\psi_s=\frac{1}{2}\,\text{erfc}\left(\frac{z_r}{\sqrt 2}\right)~~.\label{psir}
\end{gather}
The distributions of population sizes ($p(n)$) and resource levels ($p(r)$) can instead be found by noting that $\udl{n}=\bar n+v_n z$ and $\udl{r}=\bar r+v_r z$, with 
\begin{gather}
\bar n=\frac{c\bar\rho_r-m}{\sigma_c^2\chi_r/\nu}~~~~~,~~~~~ v_n=\frac{\nu\sqrt{q_r}}{\sigma_c\chi_r}~~,\\
\bar r=\frac{aK-c\bar\rho_n}{a+\sigma_c^2\chi_n}~~~~~,~~~~~ v_r=\frac{\sqrt{\sigma^2_c q_n+a^2\sigma^2_K}}{a+\sigma_c^2\chi_n}~~.
\end{gather}
This implies that
\begin{gather}\label{distr_th}
p(n)=\mathcal{G}(\bar n,v_n^2)\avg{\Theta(z-z_n)}_z+\delta(n)\avg{\theta(z_n-z)}_z~~,\\
p(r)=\mathcal{G}(\bar r,v^2_r)\avg{\Theta(z-z_r)}_z+\delta(r)\avg{\theta(z_r-z)}_z~~,
\end{gather}
where $\mathcal{G}(x,y)$ denotes the normal distribution with mean $x$ and variance $y$ (see Fig. \ref{distr}; notice that  abundance distributions quantitatively different from truncated Gaussians emerge when factors that are absent in our model, like space and dispersal, are explicitly accounted for; we refer the  reader to \cite{etienne} for a specific discussion of some of these aspects).
In the following we shall explore the solutions obtained with different choices for the various parameters. First, we shall set $\sigma_K=0$ (i.e., $K^\mu=K$ for all $\mu$) to focus on the role of metabolic diversity. Next we shall consider the case of heterogeneous resources ($\sigma_K>0$). In all cases, $a=1$ and $m=1$ for simplicity.

%Another advantage of the dynamical formulation is that it allows to treat quenched randomness in parameters other than the preferences $c_i^\mu$ (i.e. randomness in $m$, $K$ or $a$) {\it a posteriori} by directly averaging  Eqs (\ref{prima}-\ref{ultima}) over the corresponding distributions. 

\begin{figure*}[t!]
 \includegraphics[width=\textwidth]{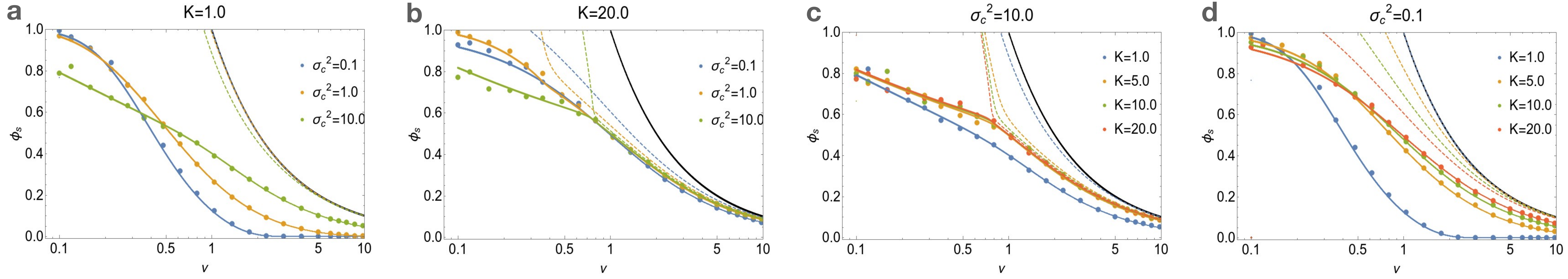}
 \caption{\label{uno}Fraction of surviving species $\phi_s$ (survival probability) as a function of $\nu$ for constant $c=1,a=1,m=1$ and fixed $K=1$ (panel a),  $K=20$ (b), $\sigma_c^2 =10$ (c) and $\sigma_c^2 =0.1$ (d) and different values of $\sigma_c^2$ or $K$ as indicated. Markers are averages over the steady states of 100 independent realizations of Eqs (\ref{ndot}) and (\ref{rdot}), continuous lines correspond to theoretical predictions, while dashed lines represent the bounds (\ref{nu}) in each case. Black line is the na\"ive bound $\phi_s=1/\nu$.} 
	\centering
	\label{uno}
    \end{figure*}  

\section{Results}

\subsection{Survival probability}

Numerical solutions of the saddle-point equations can be directly compared with results from computer simulations of (\ref{ndot},\ref{rdot}). Fig. \ref{uno}a-d shows how the fraction of surviving species varies with $\nu=N/M$ for different choices of $\sigma_c^2$ and $K$, with all other parameters fixed. Expectedly, $\phi_s$ generically decreases as the species-to-resources ratio $\nu$ increases and the ecosystem gets more competitive. The bound to $\phi_s$ given in (\ref{nu}) is however more efficiently saturated for larger values of $\nu$, $K$ and $\sigma_c$. In other terms, for any given maximal resource capacity, a higher metabolic diversity allows for a more efficient packing of species into the ecosystem. (The na\"ive bound $\phi_s=1/\nu$, where the number of surviving species equals that of resources, is also shown for comparison.) 

Notice that metabolic diversity ($\sigma_c^2$) appears to impact $\phi_s$ differently at high and low $\nu$, as a higher $\sigma_c$ seems to confer higher survival probability only in more competitive ecosystems. This can also be seen in Fig. \ref{fig:figphis}a.
	\begin{figure}
	  \includegraphics[width=0.45\textwidth]{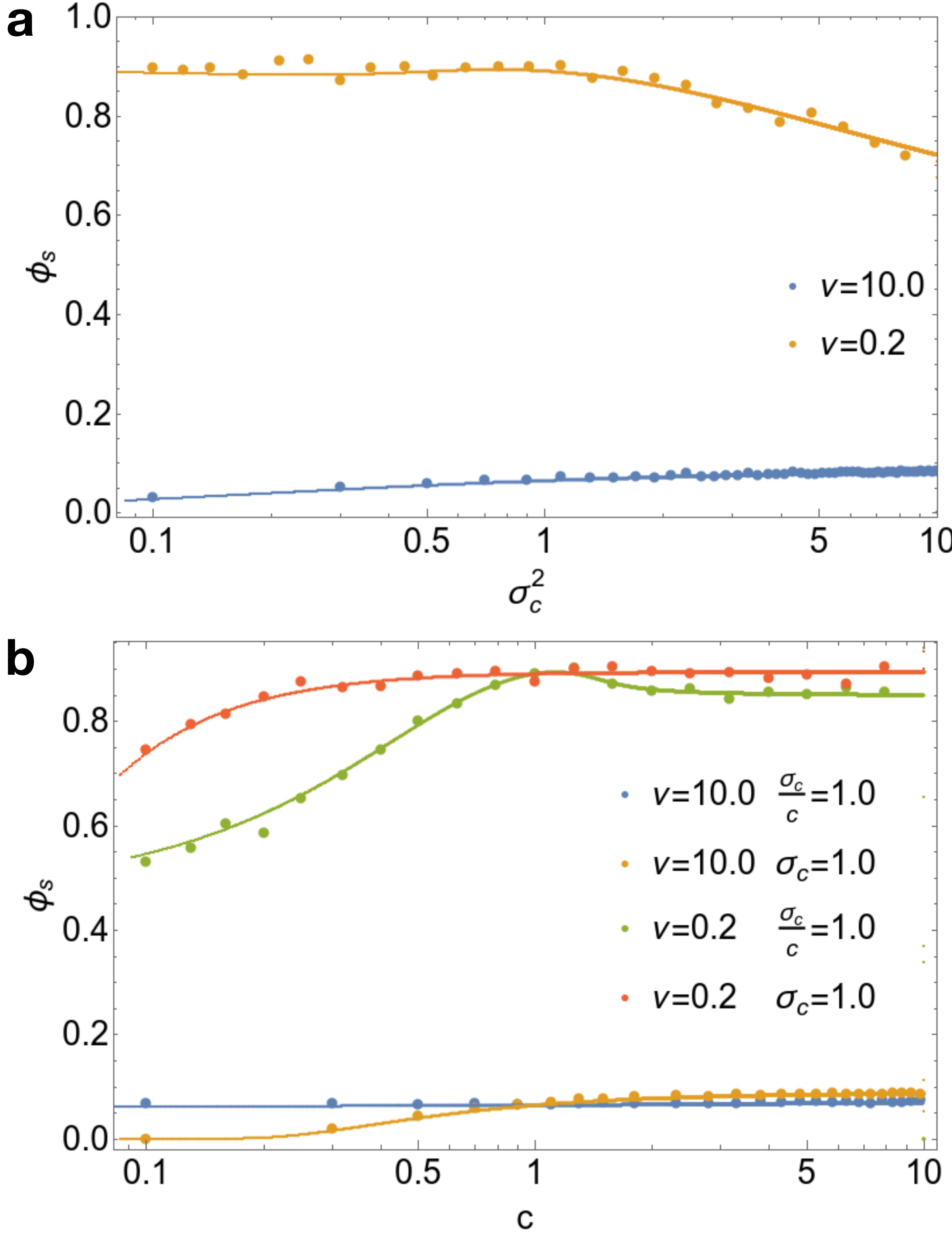}
          \caption{\label{fig:figphis} Fraction of surviving species $\phi_s$ for different values of $\nu$ as a function of $\sigma^2_c$ with $c=1$ (panel a) and as a function of $c$ for different choices of $\sigma_c$ (panel b). In both panels, $K=5,a=1,m=1$. Markers are averages over the steady states of 100 independent realizations of Eqs (\ref{ndot}) and (\ref{rdot}), while continuous lines correspond to theoretical predictions.}
	  \centering
	\label{fig:figphis}
        \end{figure}
In less competitive ecosystems (smaller $\nu$), the survival probability decreases as $\sigma_c$ increases due to the stronger impact of inefficient metabolic strategies (for any fixed $c$, larger values of $\sigma_c$ imply that negative values of $c_i^\mu$ are sampled more and more often). In other terms, the dynamics for small $\nu$ favors species that can receive a (possibly small) fitness benefit from all resources (low values of $\sigma$) over ones that receive a (possibly large) fitness benefit only from a subset of resources (higher $\sigma$). In more competitive systems (high $\nu$) such a trend is reversed, indicating that the ability to extract higher benefits from some resources provides a competitive advantage despite the costs imposed by metabolic inefficiencies (negative values of $c_i^\mu$). 

Based on these results, one expects that $\phi_s$ will increase upon increasing $c$ at fixed $\sigma_c^2$ at low $\nu$, i.e. when metabolic strategies become more efficient but more similar. This is indeed the case, as shown in Fig. \ref{fig:figphis}b (red markers). When however the increase in $c$ is accompanied by an increase of $\sigma_c$, so that $\sigma_c/c$ remains fixed (implying that strategies become more diverse as they get more efficient on average), one observes an increase of $\phi_s$ followed by a decrease at larger $c$ (Fig. \ref{fig:figphis}b, green markers). In other words, in less competitive scenarios and for any fixed relative variability of metabolic strategies, there exists a value of the average efficiency that maximizes the survival probability. Increases in efficiency at fixed $\sigma_c/c$ therefore have opposite effects at low and high $c$. In the latter regime (metabolic strategies very efficient on average), the detrimental effect of larger diversity dominates the behaviour of $\phi_s$. By contrast, when metabolic efficiency is small, improvements in this specific quantity are the key determinants of $\phi_s$. In competitive ecosystems (large $\nu$), instead, an increase in efficiency yields a survival advantage at fixed diversity $\sigma_c$, so that species can improve their survival probability by either increasing diversity (Fig. \ref{fig:figphis}a) or increasing efficiency at fixed $\sigma_c$ (Fig. \ref{fig:figphis}b, orange markers). When $\sigma_c/c$ is kept fixed, though, increased efficiency does not substantially improve $\phi_s$ (Fig. \ref{fig:figphis}b, blue markers).

	\begin{figure}
 \includegraphics[width=0.49\textwidth]{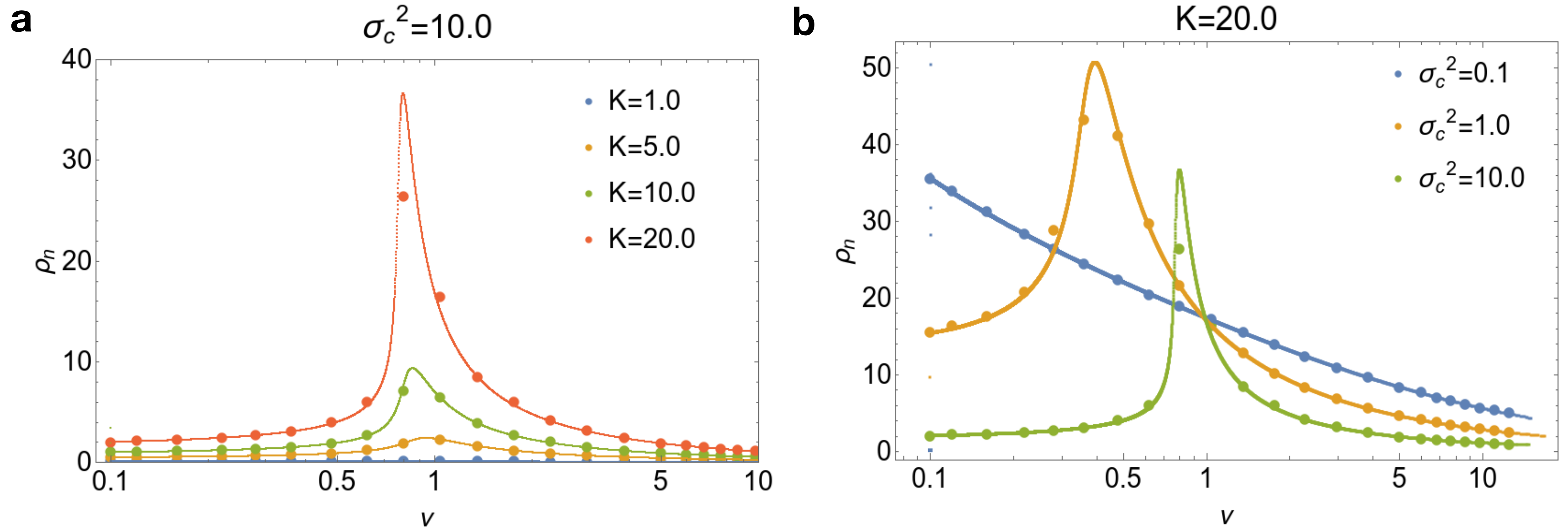}
	\caption{\label{fig:PhiN} Mean population size $\rho_n$ as a function of $\nu$ for different values of $K$ and $\sigma_c^2=10.0$ (panel a), and different values of $\sigma_c^2$ and $K=20$ (panel b). In both figures $c=1,a=1,m=1$. Markers are averages over the steady states of 100 independent realizations of Eqs (\ref{ndot}) and (\ref{rdot}), while continuous lines correspond to theoretical predictions.} 
	\centering
	\label{fig:PhiN}
        \end{figure}

\subsection{Population sizes}

The more efficient species packing found at large $\nu$ however comes at a cost for populations. Mean population sizes indeed  decrease as $\nu$ increases when packing is near optimal (Fig. \ref{fig:PhiN}a-b). This implies that, in such a regime, the introduction of new species feeds back negatively on the typical  species population size. By contrast, in less competitive ecosystems (smaller $\nu$) surviving species benefit from the introduction of new species, as their mean population sizes increase with $\nu$. Populations sizes also generically increase with increased resource capacity (larger $K$, Fig. \ref{fig:PhiN}a), while increased heterogeneity tends to reduce population sizes both in the less competitive regime (low $\nu$) and when competition is stronger and survival probability is lowest (Fig. \ref{fig:PhiN}b). This contrasts with the fact that higher $\sigma_c$ yields a positive impact (albeit weak) on the survival probability in a crowded ecosystem. To summarize, in large random instances of MacArthur's model, population sizes benefit from metabolic innovation (i.e. from the introduction of new species) when competition is weaker, while they benefit from the discovery of new resources (i.e. from a decrease in $\nu$) when the competition is strongest (larger $\nu$).

\begin{figure}[t!]
\centering
 \includegraphics[width=0.49\textwidth]{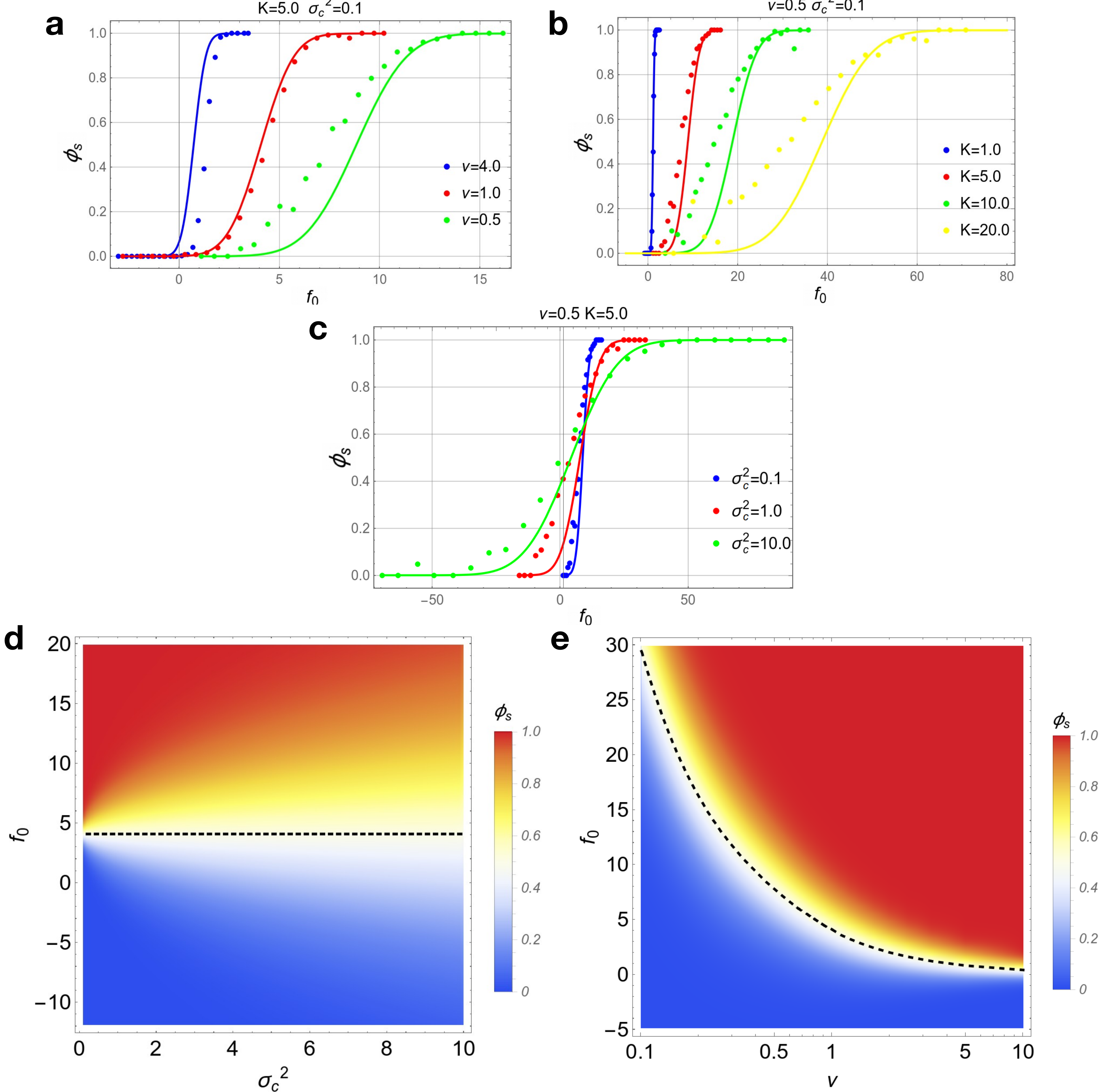}
	\caption{\label{quattro}Panels a-c) Survival probability $\phi_s$ as a function of $f_0$ for different values of $\nu$ (panel a), $K$ (panel b) and $\sigma_c^2$ (panel c), with remaining parameters fixed as indicated. d) Survival probability $\phi_s$ as a function of $f_0$ and $\sigma^2_c$ for fixed $K=5,c=1,a=1,m=1,\nu=1$. The dashed line at $f_0=4$ corresponds to $\phi_s=1/2$. e) Survival probability $\phi_s$ as a function of $f_0$ and $\nu$ for fixed $K=5,c=1,a=1,m=1,\sigma^2_c=1$ . The dashed line corresponds to $\phi_s=1/2$. }
	\label{fig:mapf0Ni}
\end{figure}

\subsection{Role of the prior fitness of species}

The dynamical approach employed for (\ref{ndot}) and (\ref{rdot}) can be easily applied to the system defined by (\ref{ndot2}) and (\ref{rdot}). The effective process now takes the form 
\begin{eqnarray}
\frac{\dot n}{n}&=& f_0-c\bar \rho_r(t)-\frac{\sigma^2_c}{\nu}\int dt'G_r(t,t')n(t')+\theta(t)+\xi_n(t)~~,\\
\frac{\dot r}{r}&=& a(K-r(t))-c\rho_n(t)-\sigma_c^2\int dt' \,G_n(t,t')\left[r(t')-K\right]+\nonumber\\
 & &+\eta(t)+\xi_r(t)~~,
\end{eqnarray}
where $f_0$ is a Gaussian random variable with mean $\frac{cK}{\nu}-m$ and variance $\frac{\sigma^2_cK^2}{\nu}$, while
\begin{gather}
\avg{\xi_n(t)\xi_n(t')}=\sigma_c^2 Q_r(t,t')~~,\\
\avg{\xi_r(t)\xi_r(t')}=\sigma_c^2 Q_r(t,t')+\sigma^2_K~~.
\end{gather}
To distinguish results for this version of the model from those of the previous sections, we shall denote the steady state quantities by an extra index $^\circ$ (e.g. $\udl{n}^\circ$ instead of $\udl{n}$). One easily sees that the equivalent of (\ref{26}) and (\ref{27}) is now 
\begin{gather}\label{nzero}
\udl{n}^\circ(z)=\frac{f_0-c\bar\rho^\circ_r+\sqrt{\sigma_c^2q_r^\circ+\sigma^2_K}z}{\sigma_c^2\chi_r^\circ/\nu}\,\Theta(z-z_n^\circ)~~,\\
\udl{r}^\circ(z)=\frac{K(a+\sigma_c^2\chi_n)-c\rho_n+\sqrt{\sigma^2q_n+\sigma^2_K}z}{a+\sigma^2_c\chi_n}\,\Theta(z-z_r^\circ)~~,\label{rzero}
\end{gather}
where 
\begin{gather}\label{znzero}
z_n^\circ=\frac{c\bar \rho^\circ_r-f_0}{\sigma_c \sqrt{q_r^\circ}}~~,\\
z_r^\circ=\frac{c\bar \rho^\circ_n-K(a+\sigma_c^2\chi^\circ_n)}{\sqrt{\sigma^2_c q_n^\circ+\sigma_K^2}}~~.
\end{gather}
In turn, saddle-point equations read
%We start the analysis, renaming the relevant magnitudes of our problem in terms of statistical properties of $f_0$. Naming the new quantites as $(\cdot )^*$, it is easy to show that they are connected to the variables discussed in the previous sections by \red{Just to be sure: the equivalent of Eqs 26 and 27 is the same but with the asterisks, right?}\blue{i guess it isn't for Eq. 26, now it must depends on $f_0$. It should be: $n(z)=\frac{f_0-c\rho_r+\sqrt{\sigma_c^2q_r+\sigma^2_K}z}{\sigma_c^2\chi_r/\nu}$. For $r(z)$ it should be the same as Eq. 27.   }
\begin{gather}
\bar\rho_n^\circ=\avg{\udl n^\circ(z)}_{z}~~,\\
\bar\rho_r^\circ=\frac{1}{\nu}\avg{K-\udl r^\circ(z)}_z~~,\\
q_n^\circ=\avg{[\udl{n}^{\circ}(z)]^2}_z~~,\\
q_r^\circ=\frac{1}{\nu}\avg{(K-[\udl{r}^\circ(z)])^2}_z~~,\label{qrzero}\\
\chi_n^\circ=\frac{1}{\sigma_c \sqrt{q_r^\circ}}\Big<\frac{\partial \udl n^\circ(z)}{\partial z}\Big>_z~~,\\
\chi_r^\circ=\frac{1}{\sigma_c \sqrt{q_n^\circ}}\Big<\frac{\partial \udl r^\circ(z)}{\partial z}\Big>_z~~.
\end{gather}
Because these changes do not alter the actual dynamics, the mean resource levels and population sizes must be the same, i.e. we must have
\begin{gather}
\avg{\udl n}=\avg{\udl n^\circ}~~,\\
\avg{\udl n^2}=\avg{[\udl{n}^\circ]^2}~~,\\
\avg{\udl r}=\avg{\udl r^\circ}~~,\\
\avg{\udl r^2}=\avg{[\udl{r}^\circ]^2}~~,
\end{gather}
so that 
\begin{gather}
\bar \rho_r^\circ=\frac{K}{\nu} -\bar \rho_r~~,\\
q_r^\circ=\frac{K^2}{\nu} -2K\bar \rho_r+q_r~~.
\end{gather}
Having this in mind, we can write the survival probability as a function of the prior fitness $f_0$ as:
\begin{gather}
%\phi_s(f_0)=\frac{1}{2}\;\text{erfc }\frac{z^*_n(f_0)}{\sqrt 2}~~,\\
\phi_s(f_0)=\frac{1}{2}\;\text{erfc }\left[\frac{1}{\sqrt 2}\frac{c\left(\frac{K}{\nu} -\bar \rho_r\right)-f_0}{\sigma_c \sqrt{\frac{ K^2}{\nu}-2K \bar \rho_r+q_r}}\right]~~.
\end{gather}
These results are compared with numerical simulations in Fig.  \ref{fig:mapf0Ni}. 
%\begin{figure}[t!]
%	\includegraphics[width=0.40\textwidth]{{/alpha1.0K5.0_heatmap_s2_f0_phi}.png}
%	\caption{Dependence of the survival probability on the parameters $f_0$ and $\sigma^2_c$ for  $K=5,c=1,a=1,m=1,\nu=1$. The enclosed regions are those for which $p(n>0)>0.99$ (Top region) and $p(n>0)<0.01$ (Bottom region) \red{Pls use colors instead of black/white, change $p(n>0)$ to $\phi_s$ and also pls add a line corresponding to $\phi_s=1/2$ and the mean $f_0$ (should be $4$). I don't understand whether they coincide. Also, if we have the colors I think we can remove the blue lines.}}
%	\centering
%	\label{fig:heatmap}
%\end{figure}
Notice that $\phi_s$ tends to become more and more step-like as the ecosystem becomes more competitive (larger $\nu$, Fig. \ref{fig:mapf0Ni}a), implying that the fate of species is determined to a greater extent by their prior  fitnesses. In less competitive cases, instead, lower values of $f_0$ are more easily overcome through dynamics. The carrying capacity plays a similar role (Fig. \ref{fig:mapf0Ni}b): as $K$ increases and resources become available at higher levels, species that would be less fit a priory can face better survival odds; on the other hand, the prior fitness becomes more relevant for the final fate of a species as carrying capacities decrease. Likewise (Fig. \ref{fig:mapf0Ni}c), larger heterogeneities in metabolic strategies improves the survival probabilities of species with lower prior fitnesses. The dependences on $\sigma_c^2$ and $\nu$ are represented more completely in the heat maps shown in Figs \ref{fig:mapf0Ni}d and \ref{fig:mapf0Ni}e.

\subsection{Stability}

\begin{figure}
 \includegraphics[width=0.48\textwidth]{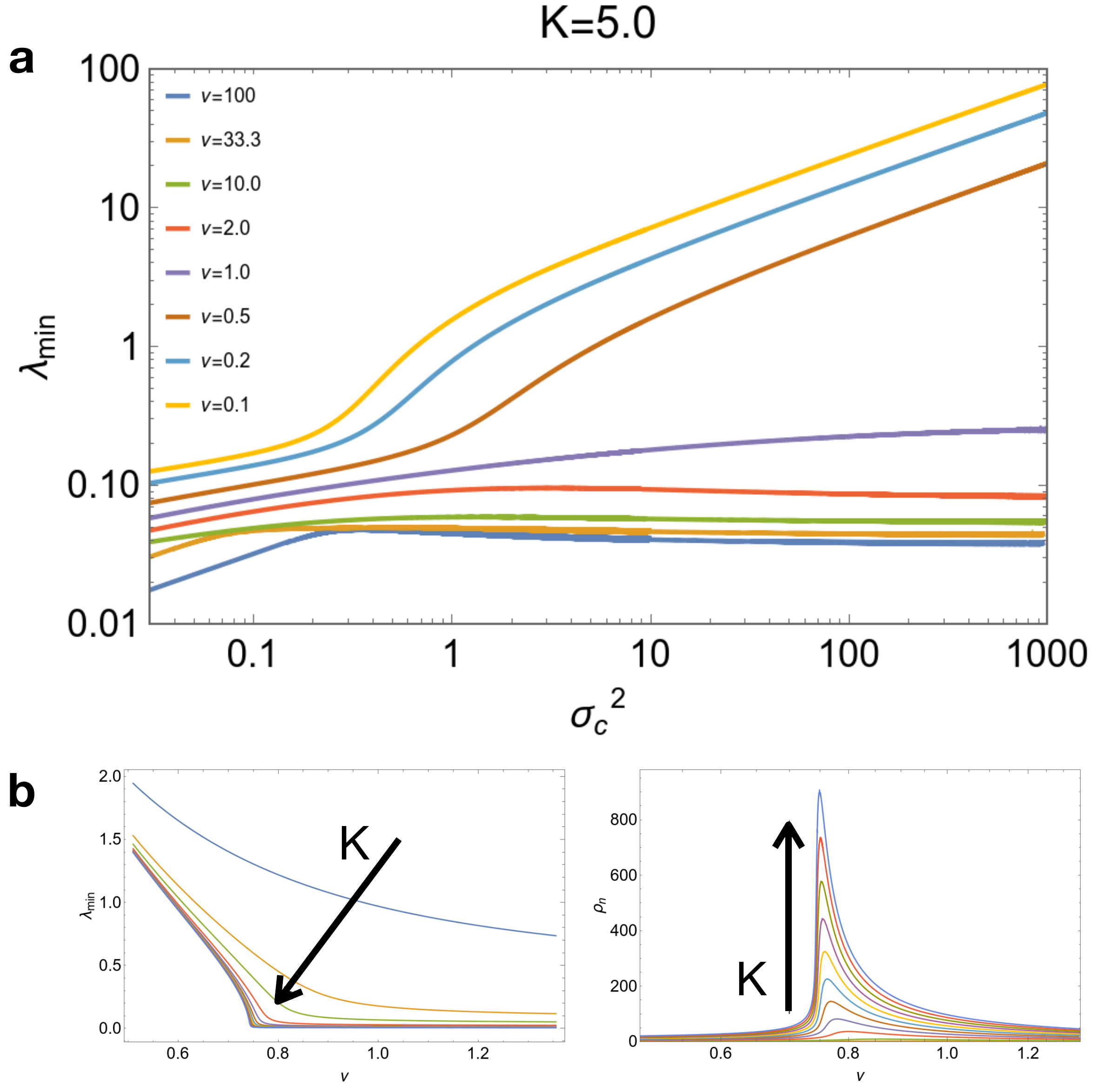}
  	\caption{\label{cinque} (a) Behaviour of $\lambda_{\min}$, Eq. (\ref{lmin}), versus $\sigma_c^2$ for fixed $K=5, a=1, m=1$ and different $\nu$. (b) Behaviour of $\lambda_{\min}$ (left) and $\rho_n$ (right) versus $\nu$ for different values of $K$ (increasing in the direction of the arrow).}
\end{figure}

Fig. \ref{cinque}a displays the eigenvalue $\lambda_{\min}$ given in (\ref{lmin}) as a function of the metabolic heterogeneity $\sigma^2_c$ and for various $\nu$, with all other parameters fixed. While this quantity strictly speaking controls stability when the characteristic timescales of species and resources are widely separated, it can still provide useful indications about how different parameters may affect the stability of MacArthur's model. We focus in particular on the metabolic heterogeneity $\sigma^2_c$. 
At low species-to-resources ratios, $\lambda_{\min}$ generically increases with $\sigma_c$, indicating that higher metabolic heterogeneity improves stability. At high $\nu$, however, i.e. in more competitive situations, $\lambda_{\min}$ displays a maximum at relatively small values of $\sigma_c^2$, indicating that the system becomes less stable when metabolic heterogeneity is too large. One therefore understands that $\lambda_{\min}$ also decreases systematically as $c$ increases at fixed $\sigma_c^2$, as strategies become more homogeneous (not shown). Likewise, one easily sees from (\ref{lmin}) that the introduction of new species (increase $\nu$) always decreases stability, in agreement with the classical results of \cite{may72}. 

Fig. \ref{cinque}b links instead the behaviour of the mean population size $\rho_n$ for increasing values of $K$ to that of $\lambda_{\min}$. In particular, the value of $\nu$ at which $\rho_n$ tends to peak for $K\to\infty$ corresponds to the point where the ecosystem becomes marginally stable ($\lambda_{\min}\to 0$) in the same limit. In addition, it shows that, for a sufficiently large carrying capacity and in presence of strong competition between species (large $\nu$), when surviving species saturate the achievable limit more efficiently, the ecosystem gets as close as possible to becoming unstable. Likewise, at the species-to-resources ratio that allows to sustain the largest overall populations the ecosystem gets closer and closer to an instability as $K$ increases. Despite the crude assumptions under which it was derived, Eq (\ref{lmin}) therefore does appear to provide key insight into the properties of the stationary state of the full system.

\subsection{Heterogeneous carrying capacities}

\begin{figure}
 \includegraphics[width=0.49\textwidth]{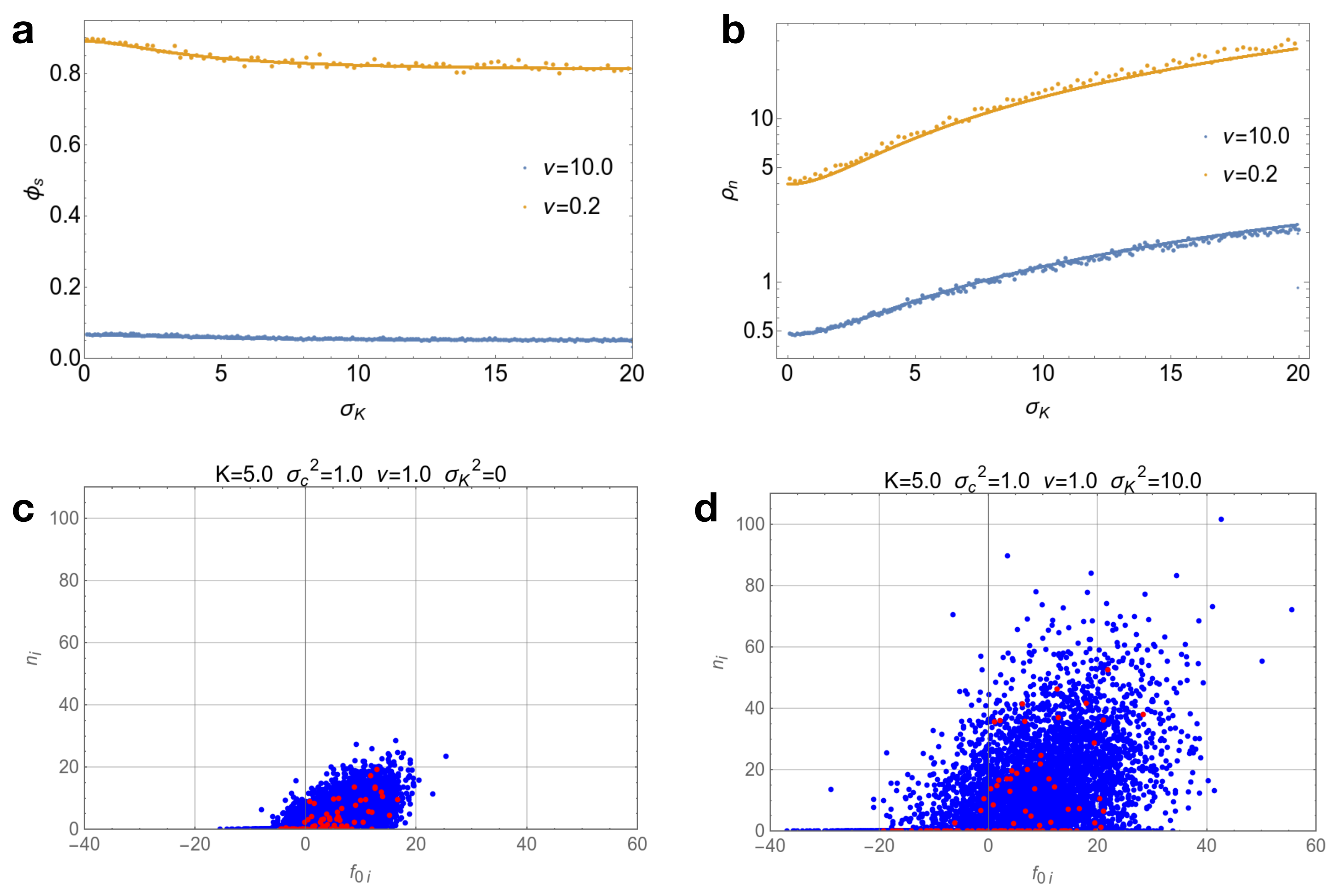}
  	\caption{\label{tre}Dependence of survival probability $\phi_s$ (panel a) and of the mean population size $\rho_n$ (panel b) on $\sigma_K$ for $\nu=10$ (orange) and $\nu=0.2$ (blue) and  $K=5.0,\sigma^2_c=a=m=1$. Markers are averages over the steady states of 100 independent realizations of Eqs (\ref{ndot}) and (\ref{rdot}), while continuous lines corresponds to theoretical predictions. (c) species population sizes ($n_i$) versus prior fitness of species ($f_{0,i}$) in the steady state of 100 realizations of Eqs (\ref{ndot}) and (\ref{rdot}), with parameters $K=5,\sigma^2_c=m=a=c=1$ and $N=100, M=100$ and $\sigma^2_K=0$. A single realization is highlighted in red. (d) Same as in (c) but with $\sigma^2_K=10$.}
  	\end{figure}

To understand the role of diversity in the availability of resources, we studied the effect of disorder in the distribution of $K^\mu$. Specifically, we considered a Gaussian distribution for $K$ with fixed mean and variance $\sigma^2_K$ changing in the interval $[0,20]$. The Gaussian distribution keeps theoretical calculations feasible while introducing negative values of $K$. The latter can be interpreted as an outflow of resources from the environment (sinks). To simplify the discussion, the rest of the parameters of the model were kept fixed at $\sigma_c=a=m=1$ and we focused on $\nu=10$ and $\nu=0.2$ to explore the emerging picture in highly competitive vs non competitive situations. Results are summarized in Fig. \ref{tre}.
The survival probability decreases as $\sigma^2_K$ increases (Fig. \ref{tre}a), as a consequence of the enhancement of resource depletion (see also Eqs (\ref{zr}) and (\ref{psir}): as $\sigma^2_K$ increases, $z_r$ and, in turn, $\psi_s$ decrease rapidly). On the other hand, mean population sizes increase markedly as $\sigma_K$ increases (Fig. \ref{tre}b). In other words, stronger variability in resource levels (including sinks) has a (small) negative impact on species survival probability but allows for the sustainment of a larger number of individuals. Both of these features follow directly from the analytical solution. According to Eq. (\ref{nzero}), the mean population level increases as $\sigma_K$ increases. On the other hand, so does the mean resource level (see Eq. (\ref{rzero})), which in turn causes $z_n^\circ$ (Eq. (\ref{znzero})) to increase as $q_r^\circ$ (Eq. (\ref{qrzero})) gets smaller. An increase of $z_n^\circ$ finally implies a worsening of the survival chances, as $\phi_s=\avg{\Theta(z-z_n^\circ)}_z$. The difference between systems with and without large variability in the level of resources is further highlighted in Figs \ref{tre}c and \ref{tre}d. One sees that, for all other parameters fixed, in presence of heterogeneities in the carrying capacities, prior fitnesses span a much broader range. However, species with the same prior fitness can achieve much larger population sizes in a heterogeneous environment than in a homogeneous one.

\section{Discussion}

MacArthur's model provides a versatile theoretical framework to study emergent properties in large complex ecosystems, with much room for improvements that bring the model closer to reality and allow for testable predictions. In analyzing the simplest version, as done here, one mostly aims at defining the set of allowed behaviours and identifying the elementary mechanisms driving the response of the ecosystem to perturbations. Both, at the level of species (e.g. the introduction of new species) or resources (e.g. the disappearance of a viable resource). Understanding how such perturbations affect global features like stability, survival probability etc. is indeed the focus point of statistical mechanics approaches. To summarize, within MacArthur's model, metabolic heterogeneity far from generically providing an advantage in terms of stability or survival, bears different effects in more versus less competitive situations. In a very competitive scenario, metabolic heterogeneity favors the survival probability, on the contrary, in a less competitive scenario, the survival probability decrease, because metabolic heterogeneity translates essentially into species with lower fitness. Likewise, heterogeneity at the level of resources leads to higher population sizes for smaller numbers of species rather than favoring higher survival probabilities. Still, diversity at steady state in this system is limited by competitive exclusion \cite{hardin60,levins64}, which in our case takes the form of Eq. (\ref{nu}). 

In our view, the key defining aspect of MacArthur's model is given by the fact that the growth rate of individual species is a linear superposition of contributions due to different resources. Such an assumption is unrealistic in many cases, as e.g. when some resources are essential and the growth rate of a species is positive only when that species has access to at least one of these resources \cite{dubinkina19} or in presence of a bound in the quantities and quality of the resources that can be consumed \cite{posfai17}. Experimental studies of microbes under co-utilization of different resources have likewise displayed a more complicated picture \cite{hermsen15}. Alternatively, other models capable of overcoming the `niche' scenario exploit dynamical effects \cite{levins79,descamps05}, and higher-order interactions have also been shown to impact ecosystem diversity in a non-trial way \cite{bairey16,grilli17}. The emergent phenomenology of consumer-resource models in such cases is markedly more complex. To our knowledge, a statistical-mechanics treatment of fully disordered cases is still lacking, and techniques like the path-integral method employed here seem to be a proper starting point for the analysis of these models in the limit of large system sizes. 

On another level, a more realistic theory would require the integration of a more detailed descriptions of metabolism into the model, possibly along with features like microbial growth laws \cite{pacciani20} and cross-feeding \cite{liao20} or considering directly how the by-products of metabolism shape the environment, leading to secondary interactions between the species \cite{cossio20}. Work along these lines is likely to shed new light about the origin, stability and reproducibility of species compositions in extended ecosystems.

\acknowledgments We are indebted with Tobias Galla and Pankaj Mehta for useful discussions. Work supported by Horizon 2020 Marie Sk{\l}odowska-Curie Action-Research and Innovation Staff Exchange (MSCA-RISE) 2016 grant agreement 734439 (INFERNET: New algorithms for inference and optimization from large-scale biological data) and partially funded by the CITMA Project of the Republic of Cuba, PNCB-Statistical Mechanics of Metabolic Interactions.

\section*{Data availability statement}

Data sharing is not applicable to this article as no new data were created or analyzed in this study.

\appendix
\numberwithin{equation}{section}
\renewcommand{\theequation}{\thesection\arabic{equation}}
\begin{widetext}

\section{Derivation of the effective processes}

We begin by re-writing the dynamics (\ref{ndot}) and (\ref{rdot}) as
  \begin{gather}\label{ndotth}
    \frac{\dot n_i}{n_i}-\sum_\mu c_i^\mu r^\mu+m_i-\theta_i=0~~,\\
    \frac{\dot r^\mu}{r^\mu}-a^\mu(K^\mu-r^\mu)+\sum_j c_j^\mu n_j-\theta^\mu=0~~,\label{rdotth}
    \end{gather}
where we added auxiliary time-dependent probing fields $\theta_i$ and $\theta^\mu$. Our goal here is to evaluate the dynamical generating function
\begin{equation}\label{dgenf}
Z[\boldsymbol{\psi,\phi}] \equiv  \ovl{\avg{e^{\ii\sum_i\int dt \psi_i(t)n_i(t)+\ii\sum_\mu\int dt \phi^\mu(t)r^\mu(t)}}_\mathrm{paths}}~~,
\end{equation}
by explicitly carrying out the averages over trajectories of (\ref{ndotth}) and (\ref{rdotth}) (`paths') and over the quenched disorder, represented by the coefficients $c_i^\mu$ and by the carrying capacities $K^\mu$. Correlations and response functions for population sizes are linked to $Z$ by
\begin{gather}
-\ii\lim_{\substack{\boldsymbol{\psi}\to 0\\ \boldsymbol{\phi}\to 0}}\frac{\partial Z}{\partial \psi_i(t)}=\ovl{\avg{n_i(t)}_\mathrm{paths}}~~,\\
-\ii\lim_{\substack{\boldsymbol{\psi}\to 0\\ \boldsymbol{\phi}\to 0}}\frac{\partial^2 Z}{\partial \psi_i(t)\partial \psi_j(t')}=\ovl{\avg{n_i(t)n_j(t')}_\mathrm{paths}}~~,\\
-\ii\lim_{\substack{\boldsymbol{\psi}\to 0\\ \boldsymbol{\phi}\to 0}}\frac{\partial^2 Z}{\partial \psi_i(t)\partial \theta_j(t')}=\frac{\partial}{\partial\theta_j(t')}\ovl{\avg{n_i(t)}_\mathrm{paths}}
\end{gather}
(with similar formulas for means, correlations and response functions of resource levels using auxiliary fields $\boldsymbol{\phi}$ and $\theta^\mu$). The average over paths can be written explicitly in terms of the equations of motion (\ref{ndotth}) and (\ref{rdotth}) using appropriate $\delta$-distributions, namely
\begin{gather}
\avg{(\cdots)}_\mathrm{paths}\nonumber=
\int D\mathbf{n}\,D\mathbf{r}\,(\cdots)\,\prod_{i,t}\delta\left(\frac{\dot n_i}{n_i}-\sum_\mu c_i^\mu r^\mu+m_i-\theta_i\right)\,\prod_{\mu,t}\delta\left(\frac{\dot r^\mu}{r^\mu}-a^\mu(K^\mu-r^\mu)+\sum_j c_j^\mu n_j-\theta^\mu\right)~~,
%&=&\ovl{\int D(\mathbf{n,r})\;\mbox{\large $\delta$}\left(\frac{\dot n_i}{n_i}- \sum_\mu c_i^\mu r^\mu+m_i-\theta_i\;,\;\frac{\dot r^\mu}{r^\mu}-a^\mu(K^\mu-r^\mu)+\sum_j c_j^\mu n_j-\theta^\mu\right)e^{\ii\sum_i\int dt \psi_i(t)n_i(t)+\ii\sum_\mu\int dt \phi^\mu(t)r^\mu(t)}}
\end{gather}
where the product over $i$ and $t$ indicates that the condition has to be imposed for all $i$ at all times (and likewise for the product over $\mu$ and $t$). In turn, the $\delta$-functions can be expressed using the Fourier representations $\delta(y)\propto\int_{-\infty}^{\infty}\exp(\ii y\hat{y})d\hat{y}$. After simple rearrangements, this yields
\begin{eqnarray}
Z[\boldsymbol{\psi,\phi}]& = & \int D(\mathbf{n,\hatt n})D(\mathbf{r,\hatt r})\,\underbrace{e^{\ii\sum_i\int dt\,\psi_i(t)n_i(t)+\ii\sum_i\int dt\,\hatt{n_i}(t)\left(\frac{\dot{n_i}}{n_i}+m_i-\theta_i\right)}}_{{\cal A}(\mathbf{n,\hatt n})} \times \nonumber\\ & &
\underbrace{e^{\ii\sum_\mu\int dt\,\phi^\mu(t)r^\mu(t)+\ii\sum_\mu\int dt\,\hatt{r^\mu}(t)\left(\frac{\dot{r^\mu}}{r^\mu}+a^\mu r^\mu-\eta^\mu\right)}}_{{\cal B}(\mathbf{r,\hatt r})}\times
\underbrace{\prod_{i,\mu}\ovl{e^{\ii c_i^\mu\int dt\left[\hatt{r^\mu}(t)n_i(t) -r^\mu(t)\hatt{n_i}(t)\right]}}\prod_{\mu}\ovl{e^{-iK^\mu\int dt a^\mu \hat{r^\mu}}}}_{\Delta(\mathbf{n,\hatt n},\mathbf{r,\hatt r})}~~,\label{rrrwww}
\end{eqnarray}
where we have separated the term $\Delta(\mathbf{n,\hatt n},\mathbf{r,\hatt r})$ that depends on the quenched random variables $c_i^\mu$ and $K^\mu$, which conveniently factorizes. Average are easily computed since the $c_i^\mu$s (resp. $K^\mu$s) are iid Gaussian random variables with distributions $\mathcal{G}\left(c/N,\sigma_c^2/N\right)$ (resp. $\mathcal{G}\left(K,\sigma_K^2\right)$). One gets
\begin{gather}
\Delta(\mathbf{n,\hatt n},\mathbf{r,\hatt r}) =
\Delta_1(\mathbf{n,\hatt n},\mathbf{r,\hatt r})\times  e^{-\ii \sum_{\mu}a^\mu K\int dt\,\hat{r^\mu}(t)}\\
\Delta_1(\mathbf{n,\hatt n},\mathbf{r,\hatt r})=  
 e^{\ii cN\int dt\left[\lambda_r(t)\rho_n(t) -\rho_r(t)\lambda_n(t)\right]}\times\,e^{-\frac{\sigma_c^2 N}{2}\int dt dt' \left[L_r(t,t')Q_n(t,t')+Q_r(t,t')L_n(t,t')\right]}\times \nonumber\\
\times\,e^{\sigma_c^2 N\int dt dt' K_r(t,t')K_n(t,t')}\times e^{-\frac{\sigma_K^2N}{2}{(a^{\mu})}^2\int dtdt'L_r(t,t')}
\end{gather}
where we defined the macroscopic quantities:
\begin{gather}
\rho_n(t)=\frac{1}{N}\sum_i n_i(t)~~~,~~~ \rho_r(t)=\frac{1}{N}\sum_\mu r^\mu(t)\nonumber\\
\lambda_n(t)=\frac{1}{N}\sum_i\hatt{n_i}(t)~~~,~~~ \lambda_r(t)=\frac{1}{N}\sum_\mu\hatt{r^\mu}(t)\nonumber\\
Q_n(t,t')=\frac{1}{N}\sum_i n_i(t)n_i(t')~~~,~~~ Q_r(t,t')=\frac{1}{N}\sum_\mu r^\mu(t)r^\mu(t')\label{order_p}\\
L_n(t,t')=\frac{1}{N}\sum_i\hatt{n_i}(t)\hatt{n_i}(t')~~~,~~~ L_r(t,t')=\frac{1}{N}\sum_\mu\hatt{r^\mu}(t)\hatt{r^\mu}(t')\nonumber\\
K_n(t,t')=\frac{1}{N}\sum_i n_i(t)\hatt{n_i}(t')~~~,~~~ K_r(t,t')=\frac{1}{N}\sum_\mu r^\mu(t)\hatt{r^\mu}(t')~~.\nonumber
\end{gather}
To simplify the notation, we shall henceforth write $n_i$ for $n_i(t)$ and $n_i'$ for $n_i(t')$ (and likewise for all time-dependent quantities). Defining the vectors $\boldsymbol{\pi}=(\boldsymbol{\rho,\lambda,Q,L,K})$ and $\hatt{\boldsymbol{\pi}}=(\boldsymbol{\hatt{\rho},\hatt{\lambda},\hatt{Q},\hatt{L},\hatt{K}})$, we can insert each of the above definitions into $Z$ again using appropriate $\delta$-functions, obtaining
\begin{eqnarray}
Z & = & \int D(\boldsymbol{\pi,\hatt{\pi}})\, e^{N\Psi(\boldsymbol{\pi,\hatt{\pi}})+N\Phi(\boldsymbol{\pi})}\times\nonumber\\
& & \times\int D(\mathbf{n,\hatt n})\,{\cal A}(\mathbf{n,\hatt n})\, e^{-\ii\sum_i\int dt\left(\hatt{\rho_n} n_i+\hatt{\lambda_n}\hatt{n_i}\right)}\,
e^{-\ii\sum_i\int dt dt'\left[\hatt{Q_n}n_i n_i'+\hatt{L_n}\hatt{n_i}\hatt{n_i'}+\hatt{K_n} n_i\hatt{n_i'}\right]}\nonumber\\
& & \times\int D(\mathbf{r,\hatt r})\,\left[{\cal B}(\mathbf{r,\hatt r}) \times  e^{-\ii \sum_{\mu}a^\mu K\int dt\hat {r^\mu}(t)}\right]\, e^{-\ii\sum_\mu\int dt\left(\hatt{\lambda_r} \hatt{r^\mu}+\hatt{\rho_r}r^\mu\right)}\,
e^{-\ii\sum_\mu\int dt dt'\left[\hatt{Q_r}r^\mu (r^\mu)'+\hatt{L_r}\hatt{r^\mu}\hatt{r^\mu}'+\hatt{K_r} r^\mu \hatt{r^\mu}'
\right]}\label{wwwrrr}
\end{eqnarray}
where
\begin{eqnarray}
\Psi(\boldsymbol{\pi,\hatt{\pi}}) & = & \ii\int dt\left[\rho_n\hatt{\rho_n}+\rho_r\hatt{\rho_r}+\lambda_n\hatt{\lambda_n}+\lambda_r\hatt{\lambda_r}\right]\nonumber\\ 
& & +\ii\int dt dt'\left[Q_n\hatt{Q_n}+L_n\hatt{L_n}+K_n\hatt{K_n}+Q_r\hatt{Q_r}+L_r\hatt{L_r}+K_r\hatt{K_r}\right]
\end{eqnarray}
\begin{eqnarray}
\Phi(\boldsymbol{\pi})\equiv\frac{1}{N}\ln\Delta_1=\ii c\int dt\left[\lambda_r\rho_n -\rho_r\lambda_n\right]-\frac{\sigma_c^2}{2}\int dt dt' \left[Q_n L_r+Q_r L_n-2 K_n K_r\right] 
 -\frac{a^2\sigma_K^2}{2}\int dt dt' L_r~~
\end{eqnarray}
Finally, upon substituting expressions for ${\cal A}(\mathbf{n,\hatt n})$ and ${\cal B}(\mathbf{r,\hatt r})$ from (\ref{rrrwww}) into (\ref{wwwrrr}), we arrive at
\begin{equation}\label{pppooo}
Z=\int D(\boldsymbol{\pi,\hatt{\pi}})\, e^{N[\Psi(\boldsymbol{\pi,\hatt{\pi}})+\Phi(\boldsymbol{\pi})+\Omega_n(\boldsymbol{\hatt{\pi}})+\Omega_r(\boldsymbol{\hatt{\pi}})]}~~,
\end{equation}
with
\begin{eqnarray}
\Omega_n(\boldsymbol{\hatt{\pi}}) & = & \frac{1}{N}\sum_i\ln\int D(n_i,\hatt{n_i}) \, e^{\ii\int dt \,\psi_i(t)n_i(t)}\,e^{\ii\int dt\hatt{n_i}(t)\left(\frac{\dot{n_i}}{n_i}+m_i-\theta_i\right)}\times \nonumber\\
& & \times e^{-\ii\int dt \left[\hatt{\rho_n} n_i+\hatt{\lambda_n}\hatt{n_i}\right]}\,e^{-\ii\int dt dt'\left[\hatt{Q_n}n_i n_i'+\hatt{L_n}\hatt{n_i}\hatt{n_i'}+\hatt{K_n} n_i\hatt{n_i'}\right]}\\
& \equiv & \ln\int D(n,\hatt{n}) \, e^{\ii\int dt \psi(t)n(t)}\,e^{\ii\int dt\,\hatt{n}(t)\left(\frac{\dot{n}}{n}+m-\theta\right)}\times \nonumber\\
& & \times e^{-\ii\int dt \left[\hatt{\rho_n} n+\hatt{\lambda_n}\hatt{n}\right]}\,e^{-\ii\int dt dt'\left[\hatt{Q_n}n n'+\hatt{L_n}\hatt{n}\hatt{n'}+\hatt{K_n} n\hatt{n'}\right]}
\end{eqnarray}
and
\begin{eqnarray}
\Omega_r(\boldsymbol{\hatt{\pi}}) & = & \frac{1}{N}\sum_\mu\ln\int D(r^\mu,\hatt{r^\mu})\,e^{\ii\int dt \,\phi^\mu(t)r^\mu(t)}\,
e^{\ii\int dt\hatt{r^\mu}(t)\left(\frac{\dot{r^\mu}}{r^\mu}-a^\mu(K^\mu-r^\mu)-\eta^\mu\right)}\times\nonumber\\
& & \times e^{-\ii\int dt \left(\hatt{\rho_r} r^\mu+\hatt{\lambda_r}\hatt{r^\mu}\right)}
\, e^{-\ii\int dt dt'\left[\hatt{Q_r} r^\mu (r^\mu)'+\hatt{L_r}\hatt{r^\mu}\hatt{r^\mu}'+\hatt{K_r} r^\mu\hatt{r^\mu}'\right]}\\
& \equiv & \frac{1}{\nu}\ln\int D(r,\hatt{r})\,e^{\ii\int dt \,\phi(t)r(t)}\,
e^{\ii\int dt\hatt{r}(t)\left(\frac{\dot{r}}{r}-a(K-r)-\eta\right)}\times\nonumber \\
& & \times e^{-\ii\int dt \left(\hatt{\rho_r} r+\hatt{\lambda_r}\hatt{r}\right)}
\, e^{-\ii\int dt dt'\left[\hatt{Q_r} r r'+\hatt{L_r}\hatt{r}\hatt{r}'+\hatt{K_r} r\hatt{r}'\right]}~~.
\end{eqnarray}
In the limit $N\to\infty$, integrals like (\ref{pppooo}) can be evaluated by saddle point integration, implying 
\begin{equation}
Z=\int D(\boldsymbol{\pi,\hatt{\pi}})\, e^{N[\Psi(\boldsymbol{\pi,\hatt{\pi}})+\Phi(\boldsymbol{\pi})+\Omega_n(\boldsymbol{\hatt{\pi}})+\Omega_r(\boldsymbol{\hatt{\pi}})]}\sim e^{N[\Psi^\star+\Phi^\star+\Omega_n^\star+\Omega_r^\star]}~~,
\end{equation}
where the star indicates that the functions are extremized. The corresponding values of the order parameters are therefore to be obtained from the saddle point conditions
\begin{gather}
\frac{\partial\Psi}{\partial\boldsymbol{\pi}}+\frac{\partial\Phi}{\partial\boldsymbol{\pi}}=0~~,\label{nohat}\\
\frac{\partial\Psi}{\partial\boldsymbol{\hatt{\pi}}}+\frac{\partial\Omega_N}{\partial\boldsymbol{\hatt{\pi}}}+\frac{\partial\Omega_R}{\partial\boldsymbol{\hatt{\pi}}}=0~~.\label{hat}
\end{gather}
Computing derivatives explicitly, (\ref{nohat}) takes the form 
\begin{gather}
\hatt{\rho_n}=-c\lambda_r~~~,~~~\hatt{\rho_r}=c\lambda_n~~~,~~~\hatt{\lambda_n}=c\rho_r~~~,~~~\hatt{\lambda_r}=-c\rho_n\nonumber\\
\ii\hatt{Q_n}=\frac{\sigma_c^2}{2}L_r~~~,~~~\ii\hatt{Q_r}=\frac{\sigma_c^2}{2}L_n~~~,~~~\ii\hatt{L_n}=\frac{\sigma_c^2}{2}Q_r~~~,~~~\ii\hatt{L_r}=\frac{a^2\sigma_K^2}{2}+\frac{\sigma_c^2}{2}Q_n\\
\ii \hatt{K_n}=-\sigma_c^2 K_r~~~,~~~\ii \hatt{K_r}=-\sigma_c^2 K_n\nonumber~~,
\end{gather}
while for (\ref{hat}) we get
\begin{gather}
\rho_n=\avg{n}_{\star,n}~~~,~~~\rho_r=\frac{1}{\nu}\avg{r}_{\star,r}~~~,~~~\lambda_n=\avg{\hatt{n}}_{\star,n}~~~,~~~\lambda_{r}=\frac{1}{\nu}\avg{\hatt{r}}_{\star,r}\nonumber\\
Q_n=\avg{n n'}_{\star,n}~~~,~~~ Q_r=\frac{1}{\nu}\avg{r r'}_{\star,r}~~~,~~~L_n=\avg{\hatt{n} \hatt{n'}}_{\star,n}~~~,~~~L_r=\frac{1}{\nu}\avg{\hatt{r} \hatt{r}'}_{\star,r}\\
K_n=\avg{n\hatt{n'}}_{\star,n}~~~,~~~K_r=\frac{1}{\nu}\avg{r \hatt{r'}}_{\star,r}~~\nonumber,
\end{gather}
where we used the shorthands
\begin{gather}
\avg{[\cdots]}_{\star,n}=\frac{\int D(n,\hatt{n}) \,[\cdots]\, e^{\ii\int dt\,\psi(t)n(t)}\,e^{\ii\int dt\,\hatt{n}(t)\left(\frac{\dot{n}}{n}+m-\theta\right)}\times e^{-\ii\int dt \left[\hatt{\rho_n} n+\hatt{\lambda_n}\hatt{n}\right]}\,e^{-\ii\int dt dt'\left[\hatt{Q_n}n n'+\hatt{L_n}\hatt{n}\hatt{n'}+\hatt{K_n} n\hatt{n'}\right]}}{\int D(n,\hatt{n}) \, e^{\ii\int dt\,\psi(t)n(t)}\,e^{\ii\int dt\,\hatt{n}(t)\left(\frac{\dot{n}}{n}+m-\theta\right)}\times e^{-\ii\int dt \left[\hatt{\rho_n} n+\hatt{\lambda_n}\hatt{n}\right]}\,e^{-\ii\int dt dt'\left[\hatt{Q_n}n n'+\hatt{L_n}\hatt{n}\hatt{n'}+\hatt{K_n} n\hatt{n'}\right]}}~~,\\
\avg{[\cdots]}_{\star,r}=\frac{\int D(r,\hatt{r})\,[\cdots]\,e^{\ii\int dt \,\phi(t)r(t)}\,
e^{\ii\int dt\,\hatt{r}(t)\left(\frac{\dot{r}}{r}-a(K-r)-\eta\right)} \times e^{-\ii\int dt \left(\hatt{\rho_r} r+\hatt{\lambda_r}\hatt{r}\right)}
\, e^{-\ii\int dt dt'\left[\hatt{Q_r} r r'+\hatt{L_r}\hatt{r}\hatt{r}'+\hatt{K_r} r\hatt{r}'\right]}}{\int D(r,\hatt{r})\,e^{\ii\int dt \,\phi(t)r(t)}\,
e^{\ii\int dt\,\hatt{r}(t)\left(\frac{\dot{r}}{r}-a(K-r)-\eta\right)} \times e^{-\ii\int dt \left(\hatt{\rho_r} r+\hatt{\lambda_r}\hatt{r}\right)}
\, e^{-\ii\int dt dt'\left[\hatt{Q_r} r r'+\hatt{L_r}\hatt{r}\hatt{r}'+\hatt{K_r} r\hatt{r}'\right]}}~~.
\end{gather}
Order parameters involving only $\widehat{n}$ or $\widehat{r}$ can be dealt with by noting, for instance, that
\begin{gather}
L_n(t,t')\equiv\frac{1}{N}\sum_i\hatt{n_i}(t)\hatt{n_i}(t')=-\frac{1}{N}\sum_i\frac{\partial^2 Z[\boldsymbol{0,0}]}{\partial\theta_i(t)\partial\theta_i(t')}=0~~,
\end{gather}
since by definition $Z[\boldsymbol{0,0}]=1$ (see (\ref{dgenf})). One therefore easily finds that $\lambda_r=L_r=0$, implying $\hatt{\rho_n}=\hatt{Q_n}=0$, and $\lambda_n=L_n=0$, implying $\hatt{\rho_r}=\hatt{Q_r}=0$. $\Omega_r$ then reduces to
\begin{eqnarray}
\Omega_r=\frac{1}{\nu}\ln\int D(r,\hatt{r})\,e^{\ii\int dt \,\phi(t)r(t)}\,
e^{\ii\int dt\,\hatt{r}(t)\left(\frac{\dot{r}}{r}-a(K-r)-\eta+c\rho_n-\ii\sigma_c^2\int dt' K_n(t,t')r(t')\right)}\times e^{-\frac{\sigma_c^2}{2}\int dt dt'\hatt{r}Q_n\hatt{r}'}\times e^{-\frac{a^2\sigma_K^2}{2}\int dt dt'\hatt{r}\hatt{r}'}~~,
\end{eqnarray}
which corresponds to the effective resources dynamics
\begin{equation}
\frac{\dot{r}(t)}{r(t)}=a(K-r(t))-c\rho_n(t)-\sigma_c^2\int dt' G_n(t,t')r(t')+\eta(t)+\xi_r(t)~~~,~~~\avg{\xi_r(t)\xi_r(t')}=\sigma_c^2 Q_n(t,t')+a^2\sigma_K^2~~,\label{eff1}
\end{equation}
where we re-defined the response function as $-\ii K_n=G_n$ consistently with the definition of $Z$. Likewise, $\Omega_n$ takes the form 
\begin{eqnarray}
\Omega_n=\ln\int D(n,\hatt{n}) \, e^{\ii\int dt \psi(t)n(t)}\,e^{\ii\int dt\hatt{n}(t)\left(\frac{\dot{n}}{n}+m-\theta-c\rho_r-\ii\sigma_c^2\int dt' K_r(t,t,')n(t')\right)}\times e^{-\frac{\sigma_c^2}{2}\int dt dt'\hatt{n}Q_r\hatt{n'}}~~,
\end{eqnarray}
leading to
\begin{equation}
\frac{\dot{n}(t)}{n(t)}=\theta(t)-m+c\rho_r(t)-\frac{\sigma_c^2}{\nu} \int dt' G_r(t,t')n(t')+\xi_n(t)~~~,~~~\avg{\xi_n(t)\xi_n(t')}=\sigma_c^2 Q_r(t,t')~~,\label{eff2}
\end{equation}
where $-\ii K_r=\frac{1}{\nu}G_r$. Eqs (\ref{eff1}) and (\ref{eff2}) are the effective single-resource and single-species processes describing the dynamics of the full system in the limit $N,M\to\infty$ for fixed $\nu=N/M$. We have therefore linked  the original (Markovian) system described by (\ref{ndotth}) and (\ref{rdotth}) to a new (non-Markovian) system involving a single effective species and a single effective resource.

\vspace{-1.3cm}

\end{widetext}


\begin{thebibliography}{99}

\bibitem{hekstra12}Hekstra, D. R., \& Leibler, S. (2012). Contingency and statistical laws in replicate microbial closed ecosystems. Cell, 149(5), 1164-1173.

\bibitem{huttenhower12}Huttenhower, C., Gevers, D., Knight, R., Abubucker, S., Badger, J. H., Chinwalla, A. T., ... \& Giglio, M. G. (2012). Structure, function and diversity of the healthy human microbiome. Nature, 486(7402), 207.

\bibitem{friedman17}Friedman, J., Higgins, L. M., \& Gore, J. (2017). Community structure follows simple assembly rules in microbial microcosms. Nature Ecology \& Evolution, 1(5), 1-7.

\bibitem{golford18}Goldford, J. E., Lu, N., Bajic, D., Estrela, S., Tikhonov, M., Sanchez-Gorostiaga, A.,  Segre, D., Mehta, P. \& Sanchez, A. (2018). Emergent simplicity in microbial community assembly. Science, 361(6401), 469-474.

\bibitem{harcombe14}Harcombe, W. R., Riehl, W. J., Dukovski, I., Granger, B. R., Betts, A., Lang, A. H., ... \& Marx, C. J. (2014). Metabolic resource allocation in individual microbes determines ecosystem interactions and spatial dynamics. Cell Reports, 7(4), 1104-1115.

\bibitem{wade16}Wade, M. J., Harmand, J., Benyahia, B., Bouchez, T., Chaillou, S., Cloez, B., ... \& Arditi, R. (2016). Perspectives in mathematical modelling for microbial ecology. Ecological Modelling, 321, 64-74.

\bibitem{succurro18}Succurro, A., \& Ebenh\"oh, O. (2018). Review and perspective on mathematical modeling of microbial ecosystems. Biochemical Society Transactions, 46(2), 403-412.

\bibitem{biroli18}Biroli, G., Bunin, G., \& Cammarota, C. (2018). Marginally stable equilibria in critical ecosystems. New Journal of Physics, 20(8), 083051.

\bibitem{goyal18}Goyal, A., \& Maslov, S. (2018). Diversity, stability, and reproducibility in stochastically assembled microbial ecosystems. Physical Review Letters, 120(15), 158102.

\bibitem{macarthur70}MacArthur, R. (1970). Species packing and competitive equilibrium for many species. Theoretical Population Biology, 1(1), 1-11.

\bibitem{chesson90}Chesson, P. (1990). MacArthur's consumer-resource model. Theoretical Population Biology, 37(1), 26-38.

\bibitem{demartino06}De Martino, A., \& Marsili, M. (2006). Statistical mechanics of socio-economic systems with heterogeneous agents. Journal of Physics A: Mathematical and General, 39(43), R465.

\bibitem{yoshino07}Yoshino, Y., Galla, T., \& Tokita, K. (2007). Statistical mechanics and stability of a model eco-system. Journal of Statistical Mechanics: Theory and Experiment, 2007(09), P09003.

\bibitem{tikhonov16}Tikhonov, M. (2016). Community-level cohesion without cooperation. Elife, 5, e15747.

\bibitem{tikhonov16-2}Tikhonov, M. (2016). Theoretical ecology without species. Bulletin of the American Physical Society, 61.

\bibitem{tikhonov17}Tikhonov, M., \& Monasson, R. (2017). Collective phase in resource competition in a highly diverse ecosystem. Physical Review Letters, 118(4), 048103.

\bibitem{advani18}Advani, M., Bunin, G., \& Mehta, P. (2018). Statistical physics of community ecology: a cavity solution to MacArthur's consumer resource model. Journal of Statistical Mechanics: Theory and Experiment, 2018(3), 033406.

\bibitem{tikhonov18}Tikhonov, M., \& Monasson, R. (2018). Innovation rather than improvement: a solvable high-dimensional model highlights the limitations of scalar fitness. Journal of Statistical Physics, 172(1), 74-104.

\bibitem{cui20}Cui, W., Marsland III, R., \& Mehta, P. (2020). Effect of resource dynamics on species packing in diverse ecosystems. Physical Review Letters, 125(4), 048101.

\bibitem{marsland20}Marsland, R., Cui, W., \& Mehta, P. (2020). A minimal model for microbial biodiversity can reproduce experimentally observed ecological patterns. Scientific Reports, 10(1), 1-17.

\bibitem{pacciani20}Pacciani-Mori, L., Giometto, A., Suweis, S., \& Maritan, A. (2020). Dynamic metabolic adaptation can promote species coexistence in competitive communities. PLoS Computational Biology, 16(5), e1007896.

\bibitem{dedo78}De Dominicis, C. (1978). Dynamics as a substitute for replicas in systems with quenched random impurities. Physical Review B, 18(9), 4913.

\bibitem{sompo82}Sompolinsky, H., \& Zippelius, A. (1982). Relaxational dynamics of the Edwards-Anderson model and the mean-field theory of spin-glasses. Physical Review B, 25(11), 6860.

\bibitem{sengupta99}Sengupta, A. M., \& Mitra, P. P. (1999). Distributions of singular values for some random matrices. Physical Review E, 60(3), 3389.

\bibitem{etienne}Etienne, R. S., \& Alonso, D. (2005). A dispersal-limited sampling theory for species and alleles. Ecology letters, 8(11), 1147.

\bibitem{may72}May, R. M. (1972). Will a large complex system be stable? Nature, 238, 413.

\bibitem{hardin60}Hardin, G. (1960). The competitive exclusion principle. Science, 131, 1292.

\bibitem{levins64}MacArthur, R., \& Levins, R. (1964). Competition, habitat selection, and character displacement in a patchy environment. Proceedings of the National Academy of Sciences of the United States of America, 51, 1207.

\bibitem{dubinkina19}Dubinkina, V., Fridman, Y., Pandey, P. P., \& Maslov, S. (2019). Multistability and regime shifts in microbial communities explained by competition for essential nutrients. eLife, 8, e49720.

\bibitem{posfai17}Posfai, A., Taillefumier, T., \& Wingreen, N. S. (2017). Metabolic trade-offs promote diversity in a model ecosystem. Physical Review Letters, 118(2), 028103.

\bibitem{hermsen15}Hermsen, R., Okano, H., You, C., Werner, N., \& Hwa, T. (2015). A growth-rate composition formula for the growth of E. coli on co-utilized carbon substrates. Molecular Systems Biology, 11(4), 801.

\bibitem{levins79}Levins, R. (1979). Coexistence in a variable environment. The American Naturalist, 114(6), 765-783.

\bibitem{descamps05}Descamps-Julien, B., \& Gonzalez, A. (2005). Stable coexistence in a fluctuating environment: an experimental demonstration. Ecology, 86(10), 2815-2824.

\bibitem{bairey16}Bairey, E., Kelsic, E. D., \& Kishony, R. (2016). High-order species interactions shape ecosystem diversity. Nature communications, 7, 1-7.

\bibitem{grilli17}Grilli, J., Barabas, G., Michalska-Smith, M. J., \& Allesina, S. (2017). Higher-order interactions stabilize dynamics in competitive network models. Nature, 548(7666), 210-213.





\bibitem{liao20}Liao, C., Wang, T., Maslov, S., \& Xavier, J. B. (2020). Modeling microbial cross-feeding at intermediate scale portrays community dynamics and species coexistence. PLoS Computational Biology, 16(8), e1008135




\bibitem{cossio20}Fernandez-de-Cossio-Diaz, J., \& Mulet, R. (2020). Statistical mechanics of interacting metabolic networks. Physical Review E, 101(4), 042401.


\end{thebibliography}
\end{document}